\documentclass[12pt]{article}
\usepackage{lscape}
\usepackage{epsfig}
\usepackage{wrapfig}
\usepackage{subfigure}
\usepackage{amsmath}
\usepackage{hhline}
\usepackage{amssymb}
\usepackage{times}
\usepackage{cite}

\newlength{\dinwidth}
\newlength{\dinmargin}
\begin{document}  
\newcommand{\sigrdarg}{\sigma_r^{D(3)}(\xpom,\beta, Q^2)}

\newcommand{\PO}{{I\!\!P}}
\newcommand{\RO}{{I\!\!R}}
\newcommand{\pom}{{I\!\!P}}
\newcommand{\reg}{{I\!\!R}}

\newcommand{\slowpi}{\pi_{\mathit{slow}}}
\newcommand{\fiidiii}{F_2^{D(3)}}
\newcommand{\fiidiiiarg}{\fiidiii\,(\beta,\,Q^2,\,x)}
\newcommand{\n}{1.19\pm 0.06 (stat.) \pm0.07 (syst.)}
\newcommand{\nz}{1.30\pm 0.08 (stat.)^{+0.08}_{-0.14} (syst.)}
\newcommand{\fiidiiiful}{F_2^{D(4)}\,(\beta,\,Q^2,\,x,\,t)}
\newcommand{\fiipom}{\tilde F_2^D}
\newcommand{\ALPHA}{1.10\pm0.03 (stat.) \pm0.04 (syst.)}
\newcommand{\ALPHAZ}{1.15\pm0.04 (stat.)^{+0.04}_{-0.07} (syst.)}
\newcommand{\fiipomarg}{\fiipom\,(\beta,\,Q^2)}
\newcommand{\pomflux}{f_{\pom / p}}
\newcommand{\nxpom}{1.19\pm 0.06 (stat.) \pm0.07 (syst.)}
\newcommand {\gapprox}
   {\raisebox{-0.7ex}{$\stackrel {\textstyle>}{\sim}$}}
\newcommand {\lapprox}
   {\raisebox{-0.7ex}{$\stackrel {\textstyle<}{\sim}$}}
\def\gsim{\,\lower.25ex\hbox{$\scriptstyle\sim$}\kern-1.30ex%
\raise 0.55ex\hbox{$\scriptstyle >$}\,}
\def\lsim{\,\lower.25ex\hbox{$\scriptstyle\sim$}\kern-1.30ex%
\raise 0.55ex\hbox{$\scriptstyle <$}\,}
\newcommand{\pomfluxarg}{f_{\pom / p}\,(x_\pom)}
\newcommand{\dsf}{\mbox{$F_2^{D(3)}$}}
\newcommand{\dsfva}{\mbox{$F_2^{D(3)}(\beta,Q^2,x_{I\!\!P})$}}
\newcommand{\dsfvb}{\mbox{$F_2^{D(3)}(\beta,Q^2,x)$}}
\newcommand{\dsfpom}{$F_2^{I\!\!P}$}
\newcommand{\gap}{\stackrel{>}{\sim}}
\newcommand{\lap}{\stackrel{<}{\sim}}
\newcommand{\fem}{$F_2^{em}$}
\newcommand{\tsnmp}{$\tilde{\sigma}_{NC}(e^{\mp})$}
\newcommand{\tsnm}{$\tilde{\sigma}_{NC}(e^-)$}
\newcommand{\tsnp}{$\tilde{\sigma}_{NC}(e^+)$}
\newcommand{\st}{$\star$}
\newcommand{\sst}{$\star \star$}
\newcommand{\ssst}{$\star \star \star$}
\newcommand{\sssst}{$\star \star \star \star$}
\newcommand{\tw}{\theta_W}
\newcommand{\sw}{\sin{\theta_W}}
\newcommand{\cw}{\cos{\theta_W}}
\newcommand{\sww}{\sin^2{\theta_W}}
\newcommand{\cww}{\cos^2{\theta_W}}
\newcommand{\trm}{m_{\perp}}
\newcommand{\trp}{p_{\perp}}
\newcommand{\trmm}{m_{\perp}^2}
\newcommand{\trpp}{p_{\perp}^2}
\newcommand{\alp}{\alpha_s}
\newcommand{\etamax}{\eta_{\rm max}}

\newcommand{\alps}{\alpha_s}
\newcommand{\sqrts}{$\sqrt{s}$}
\newcommand{\LO}{$O(\alpha_s^0)$}
\newcommand{\Oa}{$O(\alpha_s)$}
\newcommand{\Oaa}{$O(\alpha_s^2)$}
\newcommand{\PT}{p_{\perp}}
\newcommand{\JPSI}{J/\psi}
\newcommand{\sh}{\hat{s}}
\newcommand{\uh}{\hat{u}}
\newcommand{\MP}{m_{J/\psi}}
\newcommand{\xbj}{x}
\newcommand{\xpom}{x_{\PO}}
\newcommand{\ttbs}{\char'134}
\newcommand{\xpomlo}{3\times10^{-4}}  
\newcommand{\xpomup}{0.05}  
\newcommand{\dgr}{^\circ}
\newcommand{\pbarnt}{\,\mbox{{\rm pb$^{-1}$}}}
\newcommand{\gev}{\,\mbox{GeV}}
\newcommand{\WBoson}{\mbox{$W$}}
\newcommand{\fbarn}{\,\mbox{{\rm fb}}}
\newcommand{\fbarnt}{\,\mbox{{\rm fb$^{-1}$}}}
\newcommand{\dsdx}[1]{$d\sigma\!/\!d #1\,$}
\newcommand{\eV}{\mbox{e\hspace{-0.08em}V}}
%
%
\newcommand{\qsq}{\ensuremath{Q^2} }
\newcommand{\gevsq}{\ensuremath{\mathrm{GeV}^2} }
\newcommand{\et}{\ensuremath{E_t^*} }
\newcommand{\rap}{\ensuremath{\eta^*} }
\newcommand{\gp}{\ensuremath{\gamma^*}p }
\newcommand{\dsiget}{\ensuremath{{\rm d}\sigma_{ep}/{\rm d}E_t^*} }
\newcommand{\dsigrap}{\ensuremath{{\rm d}\sigma_{ep}/{\rm d}\eta^*} }

\newcommand{\dstar}{\ensuremath{D^*}}
\newcommand{\dstarp}{\ensuremath{D^{*+}}}
\newcommand{\dstarm}{\ensuremath{D^{*-}}}
\newcommand{\dstarpm}{\ensuremath{D^{*\pm}}}
\newcommand{\zDs}{\ensuremath{z(\dstar )}}
\newcommand{\Wgp}{\ensuremath{W_{\gamma p}}}
\newcommand{\ptds}{\ensuremath{p_t(\dstar )}}
\newcommand{\etads}{\ensuremath{\eta(\dstar )}}
\newcommand{\ptj}{\ensuremath{p_t(\mbox{jet})}}
\newcommand{\ptjn}[1]{\ensuremath{p_t(\mbox{jet$_{#1}$})}}
\newcommand{\etaj}{\ensuremath{\eta(\mbox{jet})}}
\newcommand{\detadsj}{\ensuremath{\eta(\dstar )\, \mbox{-}\, \etaj}}

\def\Journal#1#2#3#4{{#1} {\bf #2} (#3) #4}
\def\NCA{\em Nuovo Cimento}
\def\NIM{\em Nucl. Instrum. Methods}
\def\NIMA{{\em Nucl. Instrum. Methods} {\bf A}}
\def\NPB{{\em Nucl. Phys.}   {\bf B}}
\def\PLB{{\em Phys. Lett.}   {\bf B}}
\def\PRL{\em Phys. Rev. Lett.}
\def\PRD{{\em Phys. Rev.}    {\bf D}}
\def\ZPC{{\em Z. Phys.}      {\bf C}}
\def\EJC{{\em Eur. Phys. J.} {\bf C}}
\def\CPC{\em Comp. Phys. Commun.}

\begin{titlepage}

\noindent

\begin{flushleft}
DESY 12-041 \hfill ISSN 0418-9833 \\
March 2012
\end{flushleft}

\vspace{2cm}
\begin{center}
\begin{Large}

{\bf Inclusive Measurement of Diffractive Deep-Inelastic Scattering at HERA 
}

\vspace{2cm}

H1 Collaboration

\end{Large}
\end{center}

\vspace{2cm}

\begin{abstract}

The diffractive process $ep \rightarrow eXY$,  where $Y$ denotes a proton or its low mass excitation with $M_Y<1.6$~GeV, is studied with the H1 experiment at HERA. 
The analysis is restricted to the phase space region of the photon virtuality $3~\le~Q^2~\le~1600$~GeV$^2$, the square of the four-momentum transfer at the proton vertex $|t|<~1.0$~GeV$^2$ and the longitudinal momentum fraction of the incident proton carried by the colourless exchange $x_{\pom}<0.05$.
Triple differential cross sections are measured as a function of  $x_{\pom}$, $Q^2$ and $\beta=x/x_{\pom}$ where $x$ is the Bjorken scaling variable.
These measurements are made after selecting diffractive events by demanding a large empty rapidity interval separating the final state hadronic systems $X$ and $Y$.
High statistics measurements covering the data taking periods 1999-2000 and 2004-2007 are combined with previously published results in order to provide a single set of diffractive cross sections from the H1 experiment using the large rapidity gap selection method.
The combined data represent a factor between three and thirty increase in statistics with respect to the previously published results.
The measurements are compared with predictions from NLO QCD calculations based on diffractive parton densities and from a dipole model.
The proton vertex factorisation hypothesis is tested.

\end{abstract}

\vspace{1.5cm}

\begin{center}
Submitted to \EJC
\end{center}

\end{titlepage}

\begin{flushleft}


F.D.~Aaron$^{5,48}$,           
C.~Alexa$^{5}$,                
V.~Andreev$^{25}$,             
S.~Backovic$^{30}$,            
A.~Baghdasaryan$^{38}$,        
S.~Baghdasaryan$^{38}$,        
E.~Barrelet$^{29}$,            
W.~Bartel$^{11}$,              
K.~Begzsuren$^{35}$,           
A.~Belousov$^{25}$
% ...


\bigskip{\it
% ...
$ ^{2}$ Vinca Institute of Nuclear Sciences, University of Belgrade,
          1100 Belgrade, Serbia \\
% ...
$ ^{5}$ National Institute for Physics and Nuclear Engineering (NIPNE) ,
          Bucharest, Romania$^{ m}$ \\
% ...
$ ^{11}$ DESY, Hamburg, Germany \\
% ...
$ ^{25}$ Lebedev Physical Institute, Moscow, Russia$^{ e}$ \\
% ...
$ ^{29}$ LPNHE, Universit\'e Pierre et Marie Curie Paris 6,
          Universit\'e Denis Diderot Paris 7, CNRS/IN2P3, Paris, France \\
 $ ^{30}$ Faculty of Science, University of Montenegro,
          Podgorica, Montenegro$^{ n}$ \\
% ...
$ ^{35}$ Institute of Physics and Technology of the Mongolian
          Academy of Sciences, Ulaanbaatar, Mongolia \\
% ...
$ ^{38}$ Yerevan Physics Institute, Yerevan, Armenia \\
% ...
$ ^{48}$ Also at Faculty of Physics, University of Bucharest,
          Bucharest, Romania \\
% ...
$ ^e$ Supported by the Deutsche Forschungsgemeinschaft \\
% ...
$ ^m$ Supported by the Romanian National Authority for Scientific Research
      under the contract PN 09370101 \\
 $ ^n$ Partially Supported by Ministry of Science of Montenegro,
      no. 05-1/3-3352 \\
}

\end{flushleft}
%

\newpage
\section{Introduction}

\begin{figure}[htbp] 
  \begin{center}
\includegraphics[height=4.5cm]{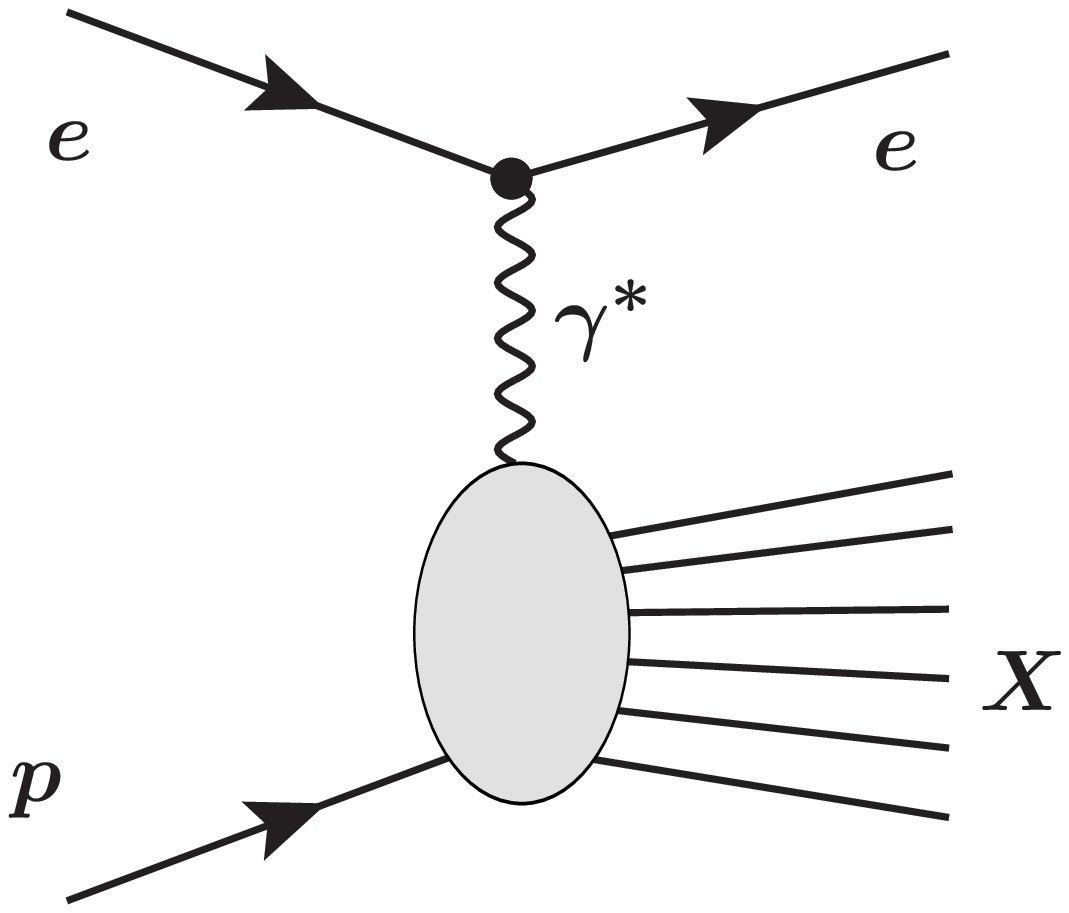}
\put(-28,-2) {{\bf (a)}}
\hspace*{1cm}
\includegraphics[height=4.5cm]{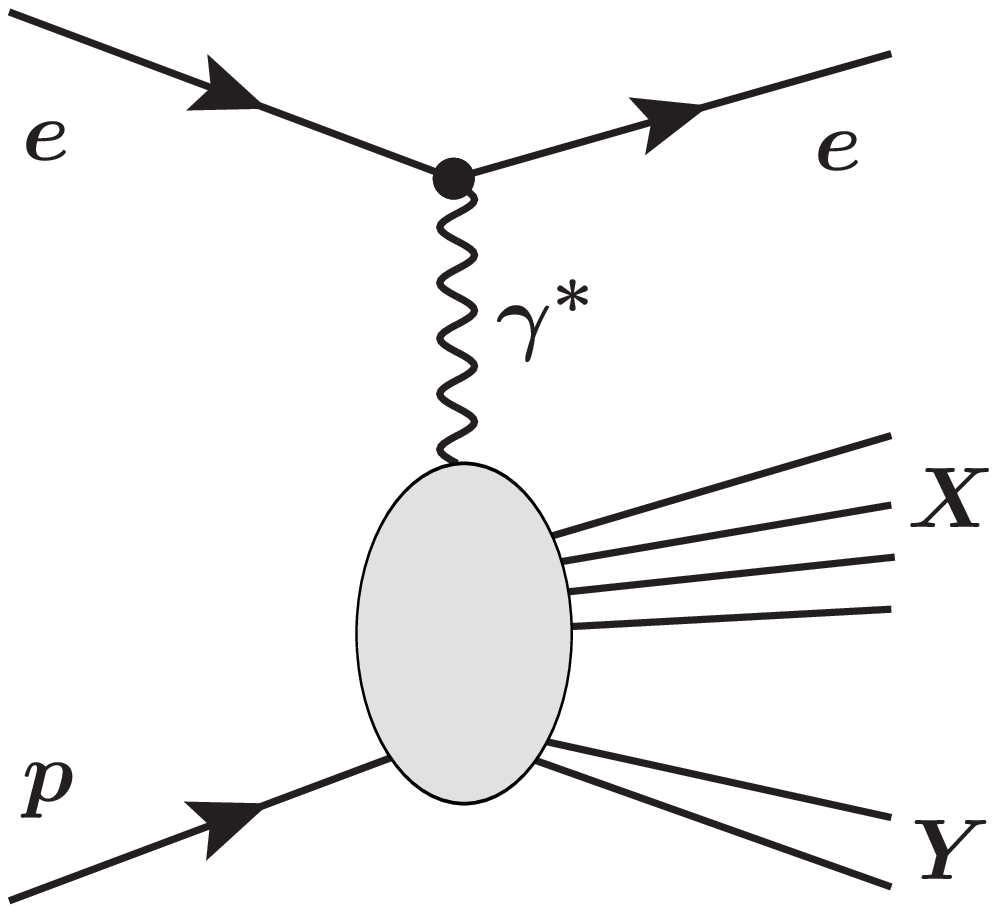}
\put(-28,-2) {{\bf (b)}}\\
  \end{center}
 \caption{Inclusive (a) and diffractive (b) deep inelastic scattering.}
 \label{fig:dis}  
 \end{figure}

At HERA a substantial fraction of up to $10\%$ of $ep$ interactions proceed via the diffractive scattering process initiated by a highly virtual photon~\cite{Derrick:1993xh,Ahmed:1994nw,h1f2d93,
h1f2d94,ZEUS:93,ZEUS:94,zeus02,zeuslps,ZEUS:97,Aktas:2006hy,Aktas:2006hx,micha,Chekanov:2008cw,Chekanov:2008fh,FLDpaper}. 
In contrast to the standard deep inelastic scattering (DIS) process $e p\to e X$ (figure~\ref{fig:dis}a), the diffractive reaction $e p\to e X Y$ contains two distinct final state systems  (figure~\ref{fig:dis}b), where $X$ is a high-mass hadronic state and $Y$ is the elastically scattered proton or its low-mass excitation, emerging from the interaction with almost the full energy of the incident proton.

The study and interpretation of diffraction at HERA provides essential inputs for the understanding of quantum chromodynamics (QCD) at high parton densities.
The sensitivity of the diffractive cross section to the gluon density at low values of Bjorken $x$ can explain the high rate of diffractive events.
Diffractive reactions may therefore be well suited to search for saturation effects in the proton structure when $x$ reaches sufficiently small values~\cite{Marquet:2007nf}.

Several theoretical QCD approaches have been proposed to interpret the dynamics of diffractive DIS.
A general theoretical framework is provided by the QCD collinear factorisation theorem for semi-inclusive DIS cross sections such as that for $ep \rightarrow eXp$ \cite{collins,Trentadue:1993ka}.  
This implies that the concept of diffractive parton distribution functions (DPDFs) may be introduced, representing conditional proton parton probability distributions under the constraint of a leading final state proton with a
particular four-momentum.  
Empirically, an additional factorisation has been found to apply to good approximation, whereby
the variables which describe the proton vertex factorise from those describing the hard interaction (proton vertex factorisation)~\cite{is,is2}. 
The dependence of the DPDFs on the kinematic variables related to the proton vertex can be parametrised conveniently using Regge formalism, which amounts to a description of diffraction in terms of the exchange of a factorisable Pomeron ($\pom$)~\cite{pomeron} with universal parton densities. 
Several authors have analysed diffractive DIS data to extract  DPDFs~\cite{h1f2d94,Aktas:2006hy,zeus:gpjets,Alvero:1998ta,Gehrmann:1995by,kgb:kw,zeuslps,ourfit,ourfit2,ourfit3,ourfit4,watt,watt2,watt3,2009qja}, with the conclusion that the data are compatible with proton vertex factorisation at low fractional proton energy losses, $\xpom$, and for photon virtualities $Q^2$ above $\sim 5$~GeV$^2$.
The DPDFs extracted  in these publications consistently find a  dominant gluon contribution.
At larger $\xpom$ ($\xpom>0.1$), a separately factorisable sub-leading Reggeon exchange ($\reg$), with a different $\xpom$ dependence and partonic composition, is usually included to maintain a good description.

The diffractive cross section can also be interpreted within the dipole model.
In this picture, the virtual photon fluctuates into a colour singlet $q {\bar q}$ pair (or dipole) of transverse size $r\!\sim\!1/Q$, which subsequently undergoes a hard scattering with the proton~\cite{dipole,dipole2,iim2,golec,Stasto:2000er,lolo}.
In the low $\beta$ domain, it is expected that $q {\bar q}$-$g$ dipoles also contribute to inclusive diffraction~\cite{GolecBiernat:1999qd}.
In a recent unified saturation description of diffractive DIS good agreement with data has been obtained in the full $Q^2$ range down to $\sim 3$~GeV$^2$~\cite{Marquet:2007nf}.
This dipole model uses the parametrisation for the dipole scattering amplitude obtained in~\cite{Soyez:2007kg}, which is an extension of the saturation model presented in~\cite{iim2} containing in addition heavy-quark contributions. 
This approach is interesting because it relates the diffractive process, in the regime $\xpom <0.01$ in which saturation is expected to be relevant, to the DIS inclusive process. 
The description of the diffractive process is obtained without extra parameter by considering the dipole cross section  $\sigma_0$ and the diffractive slope $B_D$ being directly related.

In this paper, a new measurement of the diffractive neutral current DIS cross section is presented. 
This is based upon H1 data for which there is an absence of hadronic activity in a large rapidity region extending close to the outgoing proton beam direction. 
The data were recorded with the H1 detector in the years 1999-2000 and 2004-2007, when HERA collided protons of $920$~GeV energy with $27.6$~GeV electrons and positrons.
The analysed data cover the low and medium $Q^2$ region from $3$ to $105$~GeV$^2$.
A combination with previous measurements obtained by H1, also using Large Rapidity Gap (LRG) events and based on low and medium $Q^2$ data from 1997 and high $Q^2$ data from 1999-2000~\cite{Aktas:2006hy}, is performed in order to provide a single set  of diffractive cross sections for $Q^2$ up to $1600$~GeV$^2$.
The results are compared with QCD calculations based on DPDFs extracted from previous H1 data~\cite{Aktas:2006hy} and with recent dipole model predictions~\cite{Marquet:2007nf}.

\section{Diffractive DIS Kinematics Variables and Observables}
\label{sec:kinematics}

The kinematics of the inclusive DIS process can be described by the Lorentz invariants
\begin{equation}
x=\frac{-q^2}{2P \cdot q} , \hspace{1.5cm} y=\frac{P \cdot q}{P \cdot k} ,
\hspace{1.5cm} Q^2=-q^2 \ , 
\end{equation}
\noindent where $P$ and $k$ are the 4-momenta of the incident proton and electron\footnote{
In this paper the term ``electron'' is used generically to refer to both electrons and positrons.} respectively and $q$ is the 4-momentum of the exchanged virtual photon. 
The kinematics of the diffractive process can be described in addition by the invariant masses $M_X$ and $M_Y$ of the systems $X$ and $Y$, and
\begin{eqnarray}
t     &=&  (P-P_Y)^2 \ ,  \nonumber \\ 
 && \nonumber \\[-10pt]
\beta &=& \frac{-q^2}{2q \cdot (P-P_Y)} \ = \ \frac{Q^2}{Q^2+M_X^2-t} \ , \nonumber \\
 && \nonumber \\
x_{\PO}&=& \frac{q \cdot (P-P_Y)}{q \cdot P} 
\ = \ \frac{Q^2+M_X^2-t}{Q^2+W^2-m_P^2} \ = \ \frac{x}{\beta} \ ,
\label{eqkin}
\end{eqnarray}
\noindent where $P_Y$ is the 4-momentum of system $Y$, $W^2=(q+P)^2$ is the squared centre of mass energy of the virtual photon-proton system and  $m_P$ is the proton mass.
The variable $x_{\pom}$ is the fractional momentum loss of the incident proton.  
The quantity $\beta$ has the form of a Bjorken variable defined with respect to the momentum $P-P_Y$ lost by the initial proton.

In analogy to the inclusive DIS cross section, the inclusive diffractive cross section integrated over $t$ for $ep \to eXY$ in the one-photon exchange approximation can be written in terms of diffractive structure functions $F_2^{D(3)}$ and $F_L^{D(3)}$ as
\begin{equation}
\frac{{\rm d}^3\sigma^{ep \rightarrow eXY}}{
{\rm d}Q^2\, {\rm d}\beta\, {\rm d}x_{\pom}} = 
\frac{4\pi\alpha_{\mathrm{em}}^2}{\beta Q^4}
\biggl[ \Big(1-y+\frac{y^2}{2}\Big) F_2^{D(3)}(\beta,Q^2,x_{\pom}) 
  - \frac{y^2}{2} F_L^{D(3)}(\beta,Q^2,x_{\pom}) 
\biggr] ,
\label{sigma-2}
\end{equation}
\noindent where $\alpha_{\mathrm{em}}=1/137$. 
The structure function $F_L^{D(3)}$ corresponds to longitudinal polarisation of the virtual photon.
The reduced diffractive cross section is defined by
\begin{eqnarray}
\sigma_{r}^{D(3)}(Q^2,\beta,x_{\pom})&=& 
\frac{\beta Q^4}{4 \pi \alpha_{em}^2} \
\frac{1}{(1-y+\frac{y^2}{2})} \ \frac{{\rm d}^3 \sigma^{ep \rightarrow 
eXY}}{{\rm d}Q^2 \, {\rm d} \beta \, {\rm d} x_{\pom}} \\
&=&F_2^{D(3)}-\frac{y^2}{1+(1-y)^2}F_L^{D(3)} \ .
\label{sigred}
\end{eqnarray}

\section{Experimental Procedure}

\subsection{H1 Detector}

A detailed description of the H1 detector can be found elsewhere~\cite{h1dect,h1dect2,spacal}.
Here, only the detector components relevant for the present analysis are briefly described. 
H1 uses a right-handed coordinate system with the $z$ axis along
the beam direction and the $+z$ or ``forward'' direction being that of the outgoing proton beam.
The polar angle $\theta$ is defined with respect to the $z$ axis and the
pseudorapidity is given by $\eta=-\ln \tan \theta /2$. 

The liquid argon (LAr) calorimeter ($4^{\rm \circ} \leq \theta \leq 154^{\rm \circ}$) is situated inside a solenoidal magnet. 
The energy resolutions for electromagnetic and hadronic showers are 
$\sigma(E)/E \simeq 11\% /$ $\sqrt{E/{\rm GeV}}$ $\oplus$ $1\%$ and $\sigma(E)/E \simeq 50\%/\sqrt{E/{\rm GeV}} \oplus 2\%$, respectively, as obtained from test beam measurements~\cite{Andrieu:1994yn,Andrieu:1993tz}.
The backward region ($153 ^{\rm \circ} < \theta < 176 ^{\rm \circ}$) is covered by  a lead scintillating fibre calorimeter, the SpaCal~\cite{spacal}, which has both electromagnetic and hadronic sections.
Its energy resolution for electromagnetic showers is $\sigma(E)/E \simeq 7.1\%/\sqrt{E/{\rm GeV}} \oplus 1\%$. %
A tracking chamber placed in front of the SpaCal, the backward drift chamber (BDC) for the period 1999-2000 and the backward proportional chamber (BPC) for the period 2004-2007, is used to identify the scattered electron and to determine its position.

The main component of the central tracking detector is the central jet chamber CJC ($20^{\rm \circ} < \theta < 160^{\rm \circ}$) which consists of two coaxial cylindrical drift chambers with wires parallel to the beam direction.
The measurement of charged particle transverse momenta is performed in a magnetic field of $1.16$~T, which is uniform over the full tracker volume.
The forward tracking detector, ($\theta < 30 \dgr$) is used to determine the vertex position for events where no CJC track is reconstructed.

The forward components of the H1 detector, used here to tag hadronic activity at large pseudorapidity ($3.5  < \eta < 7$), are the Plug forward calorimeter, the forward muon detector (FMD), the proton remnant tagger (PRT) and the forward tagging system (FTS).
The Plug enables energy measurements to be made in the pseudorapidity range $3.5 < \eta < 5.5$. 
It is positioned around the beam-pipe at $z = 4.9$~m. 
The FMD consists of a series of drift chambers covering the range $1.9<\eta<3.7$. 
Primary particles produced at larger $\eta$ can be detected  indirectly in the FMD if  they undergo a secondary scattering with the beam pipe or other adjacent material. 
For the period 1999-2000, secondary particles, or the scattered proton at very high $|t|$, can also be detected by the PRT, covering the range $6.5 < \eta < 7.5$, which is located at $24$~m from the interaction point and consists of layers of scintillator surrounding the beam pipe. 
In the period  2004-2007, the PRT is replaced by the FTS which consists of four stations of scintillators arranged around the proton beam pipe at $z = 26$~m, $z = 28$~m, $z = 53$~m and $z = 92$~m. 
Only the stations at $26$~m and $28$~m are used to tag proton dissociation, since further downstream elastically scattered protons often hit the beam-pipe.

The luminosity is determined from the rate of Bethe-Heitler processes measured using a calorimeter located close to the beam pipe at $z=-103$~m in the backward direction.

\subsection{Data Samples }

Different event samples corresponding to different $Q^2$ ranges are analysed in this paper.  
For the interval $3 \le Q^2 \le 25 \rm\ GeV^2$, a `minimum bias' (MB) sample corresponding to an 
integrated luminosity of $3.5 \ {\rm pb^{-1}}$ is used, which was recorded during a special data taking period in 1999 with dedicated low $Q^2$ electron triggers. 
For  photon virtualities in the interval $10 \le Q^2 \le 105 \ {\rm GeV^2}$, data taken throughout the periods 1999-2000 and 2004-2007 are used, corresponding to a total integrated luminosity of $371 \ {\rm pb^{-1}}$. 
These cross section measurements are combined with previously published H1 LRG data~\cite{Aktas:2006hy}.
All event samples are summarised in table~\ref{tab:datasets}. 

\begin{table}[h]
\centering
\begin{tabular}{|l|r@{$Q^2$}l|c|r@{$.$}l|}
\hline 
\multicolumn{1}{|c|}{Data Set} & \multicolumn{2}{|c|}{$Q^2$ range} &  \multicolumn{1}{c|}{Proton Energy}  & \multicolumn{2}{c|}{Luminosity}  \\
\multicolumn{1}{|c|}{}  & \multicolumn{2}{|c|}{(GeV$^2$)} & \multicolumn{1}{c|}{$E_p$~ (GeV)} & \multicolumn{2}{c|}{(pb$^{-1}$)}  \\
\hline 
\hline 
\multicolumn{6}{|c|}{New data samples}  \\
\hline
1999~MB       & $3 <$ \ & \ $< 25$ & $920$   & $3$ & $5$ \\
1999-2000   & $10 <$ \ & \ $< 105$ & $920$ & $34$ & $3$ \\
2004-2007   & $10 <$ \ & \ $< 105$ & $920$ & ~~~~$336$ & $6$ \\
\hline
\multicolumn{6}{|c|}{Previously published data samples}  \\
\hline
1997~MB      & $3 <$ \ & \ $< 13.5$   &  $820$ & $2$ & $0$ \\
1997         & $13.5 <$ \ & \ $< 105$ &  $820$ & $10$ & $6$ \\
1999-2000  & $133 <$ \ & \ $< 1600$ &  $920$ & $61$ & $6$ \\
\hline
\end{tabular} 
\caption{Summary of the data samples used in the analysis.}
\label{tab:datasets}
\end{table}

\subsection{Event Selection and Kinematic Reconstruction}
\label{selection}

DIS events are selected by requiring a localised energy deposit (cluster) in the SpaCal calorimeter with an energy greater than $10$~GeV, ensuring a trigger efficiency close to $100\%$.
The cluster radius of the electron candidate is required to be less than $4$~cm, as expected for an electromagnetic shower.
In order to avoid losses of energy into the beam-pipe, the radial distance between the beam axis and the cluster barycentre is required to be larger than $11$~cm.
For the data recorded in 1999-2000, a track segment was required in the BDC, matching the cluster in the SpaCal calorimeter within $3$~cm. 

Cosmic ray and beam induced backgrounds are reduced to negligible levels by requiring a vertex reconstructed within $35$~cm of 
the nominal interaction point and the timing of the signals from the tracking detector to be within the interval expected for $ep$ collisions.
Radiative events and photoproduction events in which a hadron is misidentified as the scattered electron are suppressed by requiring $\sum_i{(E^i-p_z^i)} > 37$~GeV, where $E^i$ and $p_z^i$ are the energy and longitudinal momentum of all detected particles $i$, including the scattered electron.

The inclusive DIS kinematic variables, $x$,  $Q^2$ and the inelasticity $y$, are reconstructed using the techniques introduced in~\cite{h1f2d94}. 
In order to optimise the resolution throughout the measured $y$ range, information is exploited from both the scattered  electron and the hadronic final state according to
\begin{equation}
  y = y_d + (y_e^2 - y_d^2) \ , \qquad
  Q^2 = \frac{4 E_e^2 
  \ ( 1 - y)}{\tan^2 (\theta^\prime_e / 2)} \ , \qquad x = \frac{Q^2}{s y} .
\end{equation}
\noindent Here, $y_e$ and $y_d$ denote the values of $y$ obtained from the scattered electron only (`electron
method') and from the angles of the electron and the hadronic final state (`double angle method'), respectively~\cite{Aktas:2006hy}. 
$E_e$ is the electron beam energy and $\theta^\prime_e$ is  the  polar angle of the scattered  electron.
In order to ensure a reasonable containment of the hadronic final state in the central detectors only events with $y > 0.04$ are selected.

A sub-sample of events where a diffractive exchange dominates is selected by
requiring that no signal is recorded above noise levels in a number of forward components of the H1 detector.
The pseudorapidity $\eta_{max}$ of the most forward energy deposit in the LAr calorimeter above a noise  threshold of $800$~MeV is required to be less than $3.3$.
At most one hit pair should be present in the first two layers of the FMD.
The energy measured in the Plug calorimeter is required to be smaller than $7$~GeV.
For the period 1999-2000, it is  required that there is no signal in the first five layers of the PRT.
For the period 2004-2007, it is  required that there are no hits in the $26$~m and $28$~m stations of the FTS.
After these selection criteria are applied, the systems $X$ and $Y$ are well separated by an LRG. 
The system $X$ is fully contained in the main part of the H1 detector and the system $Y$ goes unobserved into the beam pipe.

The large rapidity gap selection yields a sample which is dominated by the elastic\footnote{Here the term ``elastic'' is used to refer to the process $ep \rightarrow  eXY$ with $Y = p$ and not to $ep \rightarrow ep$.} process $ep \rightarrow eXp$, with the outgoing proton transverse momentum $p_{t,p}$, and hence $|t| \simeq p_{t,p}^2$, being relatively small. 
However, there is an admixture of proton dissociative events, $ep \rightarrow eXY$, where the proton dissociation system has a small mass $M_Y$. 
The ranges of sensitivity of the measurement in $M_Y$ and $t$ are determined by the acceptances of the forward detectors which are used to identify the large rapidity gap.
In order to keep the uncertainties arising from proton dissociation small and to ease comparisons with 
previous data~\cite{Aktas:2006hy}, the measurement is integrated over the region
$
M_Y < 1.6 \ {\rm GeV}$,
$ 
|t| < 1 \ {\rm GeV^2}.
$
The correction factors applied to account for the net migrations about these limits are determined by evaluation of the
forward detector response to elastic proton and proton dissociative processes, using the Monte Carlo program DIFFVM~\cite{diffvm}.
This correction is $9\%$ for the 1999 MB and 1999-2000 samples and $13\%$ in 2004-2007.
Noise in the forward detector components results in some events being wrongly rejected from the samples. 
These losses are determined using randomly triggered events which are overlaid with simulated events.
The reconstruction of hadrons combines information from the calorimeters and vertex-fitted tracks in the central tracker without double counting~\cite{hfsalgo}.
The reconstructed hadronic final state four vector $P_H$ is then defined as the vector sum of all reconstructed hadrons. 
The invariant mass $M_X$ of the final state system $X$ is obtained by
\begin{eqnarray}
M_X= \sqrt {P_H^2 \; \frac{y}{y_h} },
\end{eqnarray}
\noindent with $y_h = \sum_h{(E^h-p_z^h)} / 2E_e$, where the sum runs over all reconstructed hadrons. 
The factor ${y}/{y_h}$ is included to improve the resolution at large $y$, where losses in the backward direction become large. 
The kinematic reconstruction method used here leads to a resolution in $M_X$ varying from $13$ to $22\%$  in the measured kinematic range. 
In this analysis, $M_X$ is required to be above $1$~GeV.
According to equation~(\ref{eqkin}) and neglecting $t$, the diffractive variables $\beta$ and $\xpom$ are obtained from:
\begin{equation}
   \beta = \frac{Q^2}{Q^2 + M_X^2 } ; \hspace{1.5cm} 
   \xpom = \frac{x}{\beta} \ .
\end{equation}

\subsection{Monte Carlo Simulations}

Corrections for detector inefficiencies and acceptance losses due to the event selection cuts  are evaluated bin-by-bin directly from the data or by using a Monte Carlo (MC) simulation of the detectors.
Corrections for migrations in the kinematic variables due to the finite resolution are determined using MC programs.
All generated MC events are passed through a detailed, GEANT~\cite{Brun:1987ma} based, simulation of the H1
detector, which takes into account the running conditions of the different data taking periods, and are subject to the same reconstruction and analysis chain as used for data.

Diffractive DIS is modelled using the RAPGAP Monte Carlo generator~\cite{rapgap}. 
The RAPGAP event generator implements the exchange of a partonic Pomeron or meson with leading order QCD matrix elements. 
The Pomeron and meson fluxes and the parton distributions used in the event simulation are based on the DPDF fit to previous H1 data (H1 $2006$ DPDF Fit B)~\cite{Aktas:2006hy}.
At low $Q^2$, H1 $2006$ DPDF Fit B undershoots the data, as observed previously~\cite{Aktas:2006hy}. 
For $Q^2 < 7$~GeV$^2$, RAPGAP is therefore reweighted by a parametrisation, depending on $Q^2$ and $\beta$,  to describe the present data. 
Higher order QCD radiation is modelled using initial and final state parton showers in the approximation of leading logarithms~\cite{meps}. 
Hadronisation is simulated using the Lund string model~\cite{lund} as implemented in JETSET~\cite{jetset}. 
QED radiative effects, including virtual loop corrections, are taken into account via an interface to the HERACLES program~\cite{heracles}. 
Migrations into the sample from the region $M_{Y} > 5$~GeV are studied by using RAPGAP in the inclusive DIS mode.
At low $M_X$, where the presence of the meson resonances $\rho$, $\omega$, $\phi$ becomes important, the DIFFVM MC~\cite{diffvm} is used in addition.
The Monte Carlo program COMPTON~\cite{compton2} is used to simulate single dissociation and inelastic Bethe-Heitler events.

Background from $ep$ interactions may arise from photoproduction events ($Q^2 \sim 0$) in which the scattered lepton signal 
is faked by a hadron detected in the SpaCal calorimeter. 
It is estimated using the PHOJET Monte Carlo model~\cite{phojet} and found to be negligible in this analysis. 
Other backgrounds, such as those due to interactions of the beams with the remaining gas in the beam pipe or with beam line elements upstream of the H1 detector, are also found to be negligible.

\subsection{Systematic Uncertainties}
\label{sec:syst}

A detailed systematic error analysis has been performed, in which the sensitivity of the measurements to variations in the efficiencies and energy scales of the detector components and to the details of the correction procedure is tested. 
The systematic error sources leading to uncertainties which are correlated between data points are determined from the agreement of the simulation with data in this analysis and are listed below.

\begin{itemize}

\item The uncertainty on the SpaCal electromagnetic energy scale is evaluated to be $0.5\%$ and $0.4\%$ for 1999-2000 and 2004-2007 data, respectively. 
The uncertainties in the relative alignment of the different detector components are reflected in possible biases in the electron polar angle measurement at the level of $0.5$~mrad and $1$~mrad for 1999-2000 and 2004-2007 data, respectively. 
  
\item The hadronic energy scale of the LAr calorimeter is known to $2\%$ for the 1999~MB sample and to $1.5\%$ for all other samples.

\item Imperfect treatment of calorimeter noise can result in a bias in the reconstruction of $M_X$. The corresponding uncertainty is evaluated by varying the amount of calorimeter energy classified as noise by $10\%$. 
This level of precision is determined by comparing the calorimeter noise subtracted in the data with that in the Monte Carlo model, which includes a simulation of noise based on randomly triggered events.
  
\item The efficiency with which the FMD registers activity when there is hadronic energy flow in its acceptance region is varied in the simulation by $5\%$ for 1999-2000 and $4\%$ for 2004-2007. For the PRT and FTS, this efficiency 
is varied by $20\%$ and $7\%$, respectively. 
The Plug energy scale is varied by $10\%$. 
These levels of uncertainty are obtained by comparison of the present data with the Monte Carlo simulation for samples in which forward detector activity is required.
  
\item The model dependences of the acceptance and migration corrections and of the background subtractions
are estimated by varying the details of the Monte Carlo simulation within the limits permitted by the present data. 
In the RAPGAP simulation of diffraction, the $\xpom$ distribution is reweighted by $(1/\xpom)^{\pm 0.05}$, the $\beta$ distribution by $\beta^{\pm 0.05}$ and $(1-\beta)^{\pm 0.05}$, the $t$ distribution by $e^{\pm t}$~\cite{micha} and the $Q^2$ distribution by $(\log Q^2)^{\pm 0.2}$. 
The reweighting in $t$ and $(1-\beta)$ are found to have a negligible effect on the measured cross sections.
For $Q^2 < 7$~GeV$^2$, an additional uncertainty on the shape of the $\beta$ distribution is introduced to account for the  poor description of the data by RAPGAP in this phase space region.
This results in  an additional uncertainty below $1\%$ on the measured cross sections. 
The normalisation of the sub-leading meson exchange in RAPGAP is varied by $\pm 25\%$ and that of the
vector meson production simulation (DIFFVM) is varied by $\pm 50\%$. 
The uncertainty in the background from high $M_Y$, as simulated by the inclusive RAPGAP MC, is taken to be $100\%$.

\item The model dependence of the bin centre corrections is estimated by comparing the results obtained using the H1 2006 DPDF Fit A and Fit B sets. 
It results in a sizeable correlated uncertainty of up to $3\%$ only at the largest $\beta$ values.

\end{itemize}

Several further uncertainties, listed below,  affect all data points in an identical manner and are thus considered as normalisation uncertainties. 

\begin{itemize}

\item The uncertainty on the factor correcting the measured cross section to the kinematic range $M_Y < 1.6$~GeV, $|t| < 1$~GeV$^2$ is $7\%$ (see section~\ref{selection}). 
The dominant contribution to this uncertainty arises from variations in the assumed ratio of proton dissociation to elastic proton cross sections in the range $0.5$ to $2.0$. 
Fluctuations of the noise level in the forward detector components are also taken into account. 
\item The normalisation uncertainty arising from the luminosity measurement is  $1.5\%$ for the 1999~MB and 1999-2000 data samples and $3.5\%$ for 2004-2007 data.

\end{itemize}

A third class of systematic errors leads to uncertainties which are considered not to be correlated between data points. 

\begin{itemize}

\item The calculated acceptance of the $\etamax$ cut depends on the modelling of the hadronic final state topology. The associated uncertainty is estimated from the effect of using an alternative model for higher order QCD processes (the colour dipole approach~\cite{Gustafson:1986db} as implemented in ARIADNE~\cite{Lonnblad:1992tz} in place of parton showers). 
This results in an uncertainty which depends to good approximation on $\xpom$ only and varies between $1.2\%$ at $\xpom = 0.0003$ and $4\%$ at $\xpom = 0.01$. 

\item The uncertainty on the trigger efficiency is $1\%$.

\item The uncertainty on radiative corrections is $1\%$.
  
\end{itemize}

The total systematic uncertainty on each data point is formed by adding the individual contributions in quadrature.
A full decomposition of the systematic errors on the measured cross sections is available elsewhere~\cite{H1web:LRGtable}.
Away from the boundaries of the kinematic region, the systematic error excluding the normalisation uncertainty ranges from $3\%$ to $9\%$ ($4\%$ to $10\%$ for 1999 MB data), with no single source of uncertainty dominating. 
These systematic uncertainties are to be compared with statistical errors of the order of $1\%$ in the intermediate $Q^2$ domain (1999-2000 and 2004-2007 data) and $5\%$ for the low $Q^2$ region (1999 MB data).
The overall normalisation uncertainties for each data set are of the order of $7$ to $8\%$.

\section{Results and Discussion}
\label{sec:results}

\subsection{Diffractive Cross Section Measurements and Combination}

The 1999 MB, 1999-2000 and 2004-2007 data samples are used to measure the reduced diffractive cross section $\sigma_{r}^{D(3)}(Q^2,\beta,x_{\pom})$.
The bins in $Q^2$, $\beta$ and $x_{\pom}$ are chosen to have a width always larger than twice the experimental resolution. 
The cross section measurements are corrected to fixed values of $Q^2$, $\beta$ and $\xpom$ for each bin using predictions from the H1 $2006$ DPDF Fit B.
These corrections are of the order of $5\%$ in average.
The details of this procedure including bin definitions are the same as for the previous H1 measurement~\cite{Aktas:2006hy}.
The measurements are quoted at the Born level after correcting for QED radiative effects. 
Radiative corrections are calculated bin by bin using the HERACLES program~\cite{heracles} interfaced to RAPGAP.
They are smaller than $5\%$ for all measured data points. 
The results are corrected to the region $M_Y <1.6$~GeV, and $|t| \leq 1$~GeV$^2$.

The new data sets of this analysis are combined with the previously published H1 measurements from the 1997 data~\cite{Aktas:2006hy} using the $\chi^2$ minimisation method developed for the combination of inclusive DIS cross sections~\cite{F2avGlazov,Aaron:2009bp,Aaron:2009wt}.
In the year 1997, the data were taken at a centre-of-mass energy of $\sqrt{s} = 300$~GeV whilst all the other data samples were taken at $\sqrt{s} = 319$~GeV. 
The 1997 measurements are therefore corrected to $\sqrt{s} = 319$~GeV using H1 $2006$ DPDF Fit B to parametrise $F_L^{D(3)}$.
This correction is always below $1\%$ in the kinematic domain covered.
The error associated to this correction is estimated by varying the $F_L^{D(3)}$ prediction from H1 $2006$ DPDF Fit B by $\pm 100\%$, 
which is conservative with respect to the direct measurement of $F_L^{D(3)}$~\cite{FLDpaper}.
The combined cross section measurements are given for $\sqrt{s} = 319$~GeV.
For $x_{\pom} = 0.03$ and for $Q^2 > 133$~GeV$^2$ in all $x_{\pom}$ bins, only cross section values measured previously~\cite{Aktas:2006hy} are available.

\begin{figure}[p]
  \center
  \includegraphics[width=.5\textwidth]{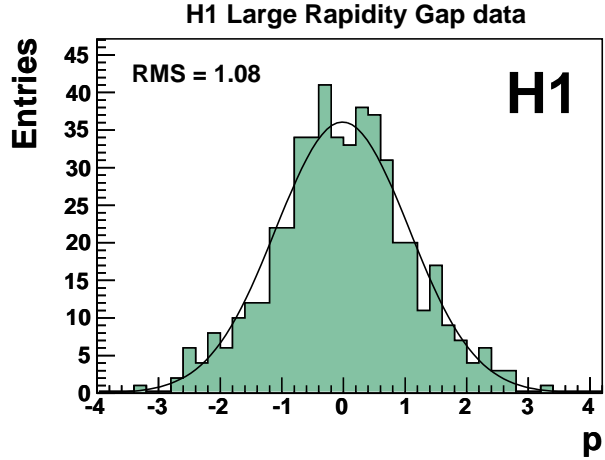}
  \caption{Distribution of pulls $p$ for all data samples. There are no entries outside the histogram range. The RMS gives the root mean square of the distribution calculated as $\overline{p^2}$. The curve shows the result of a binned log-likelihood Gaussian fit to the distribution.}
  \label{fig:pulls}
\end{figure}

The combination is performed taking into account correlated systematic uncertainties.
Systematic uncertainties associated with detector modelling are treated as uncorrelated between data sets.
Model systematic uncertainties on the acceptance and migration corrections are considered to be completely correlated between data sets.
An overall normalisation uncertainty of $4\%$ is also considered as correlated between data sets. 
It corresponds to the fraction of the correction factor accounting for smearing about the $M_Y$ and $t$ boundaries (see section~\ref{selection}), whose determination method is common to all data sets.
There are $597$ data points averaged to $277$ cross section measurements. 
The data show a reasonable consistency, with the total $\chi^2$ per degree of freedom ($n_{{\rm dof}}$) of $\chi^2/n_{{\rm dof}} = 371/320$.
The adjustments of the relative normalisations are small, with the normalisation of the 1999~MB data set staying constant and the other data samples shifting by at most $1.3\%$.
The distribution of pulls~\cite{Aaron:2009wt} of each data point relative to the combined cross section measurements is shown in figure~\ref{fig:pulls} and does not exhibit large tensions.
The largest deviations are observed in the lowest $Q^2$ bins at $x_{\pom} = 0.01$.

The $\beta$ dependence of the combined reduced cross section measurements, multiplied by $x_{\pom}$, is shown in figures~\ref{fig:beta_dep1} to~\ref{fig:beta_dep4} for fixed values of $x_{\pom}=0.0003$, $0.001$, $0.003$ and $0.01$ and is compared with the previously published cross section measurements~\cite{Aktas:2006hy} and with the prediction from the H1 $2006$ DPDF Fit B. 
The $Q^2$ dependence is presented in figure~\ref{fig:q2_dep}.
A significant reduction of statistical errors is observed.
The new combined data have a total uncertainty between $4\%$ and $7\%$ whereas they were typically of the order of $7\%$ and $10\%$ in the previously published results.

For $x_{\pom}=0.03$ only the previous measurements~\cite{Aktas:2006hy} exist. They are only slightly modified by the combination procedure. 
The resulting $\beta$ and $Q^2$ dependences are shown in figure~\ref{fig:beta_dep5}.
The results for all $\xpom$ bins are also provided in numerical form in tables~\ref{tab:data} to~\ref{tab:data_end} and in~\cite{H1web:LRGtable}.
Statistical together with uncorrelated and point-to-point correlated systematic uncertainties are shown. 
%

\subsection{Comparisons with other measurements}
\label{sec:FPS_ZEUS}

The combined reduced cross section $\sigma_r^{D(3)}$ can be compared with other H1 measurements obtained by a direct measurement of the outgoing proton using the H1 Forward Proton Spectrometer (FPS)~\cite{micha}.
The cross section $ep \rightarrow eXY$ measured here with the LRG data includes proton dissociation to any system $Y$ with a mass in the range $M_Y < 1.6$~GeV, whereas in the cross section measured with the FPS the system $Y$ is defined to be a proton.
Since the LRG and FPS data sets are statistically independent to a large extent and the dominant sources of systematic errors are different, correlations between the uncertainties on the FPS and LRG data are neglected.
The FPS results are interpolated to the $Q^2$, $\beta$ and $x_{\pom}$ bin centre values of the LRG data using a parametrisation of the H1 $2006$ DPDF Fit B. Only FPS data with interpolation corrections between $0.8$ and $1.25$ are used.
The ratio of the two measurements is then formed for each $(Q^2, \beta, x_{\pom})$ point for $\xpom=0.01$ and $\xpom = 0.03$, at which both LRG  and FPS data are available.
The global weighted average of the cross section ratio  LRG/FPS is
\begin{equation}
\frac{\sigma\left(M_Y < 1.6 \, {\rm GeV}\right)}{\sigma\left(Y=p \right)} = 1.203 \pm 0.019({\rm exp.}) \pm 0.087({\rm norm.}) \, ,
\label{eq:LRG_over_FPS}
\end{equation}
\noindent where the experimental uncertainty is a combination of statistical and uncorrelated systematic uncertainties on the measurements.
In figure~\ref{fig:FPS} the combined LRG cross section measurements as a function of $Q^2$ are compared with the interpolated FPS data rescaled by a factor $1.2$, following the above determination. 
A good agreement between the two measurements is observed.

The combined H1 LRG cross section are also compared with the most recent measurements by the ZEUS experiment using a similar LRG selection~\cite{Chekanov:2008fh}.
These ZEUS diffractive data have been determined for identical $\beta$ and $\xpom$ values, but at different $Q^2$ values to H1. 
In order to match the $M_Y < 1.6$~GeV range of the H1 data, a global factor of $0.91\pm 0.07$~\cite{Chekanov:2008fh} is applied to the ZEUS LRG data. 
The comparison for $M_Y < 1.6$~GeV between the H1 data and the rescaled ZEUS data is shown in figure~\ref{fig:ZEUS}.
The ZEUS data tend to remain higher than those of H1 by $\sim 10\%$ on average.
This difference in normalisation is consistent with the $8\%$ uncertainty on the proton-dissociation correction factor of $0.91\pm 0.07$ applied to ZEUS data combined with the normalisation uncertainties of the two data sets of $4\%$ (H1) and $2.25\%$ (ZEUS).
This normalisation difference is also similar to that of $0.85$ $\pm$ $0.01$(stat.) $\pm$ $0.03$(sys.) $^{+0.09}_{-0.12}$(norm.) between the H1 FPS and the ZEUS LPS tagged-proton data sets~\cite{micha}. 
Deviations are observed between the $\beta$ dependences of the two measurements at the highest and lowest $\beta$ values.
However a good agreement of the $Q^2$ dependence is observed throughout most of the phase space.

\subsection{Comparison with Models}
\label{sec:models}

\begin{figure}[!htbp]
\begin{center}
 \includegraphics[width=.7\textwidth]{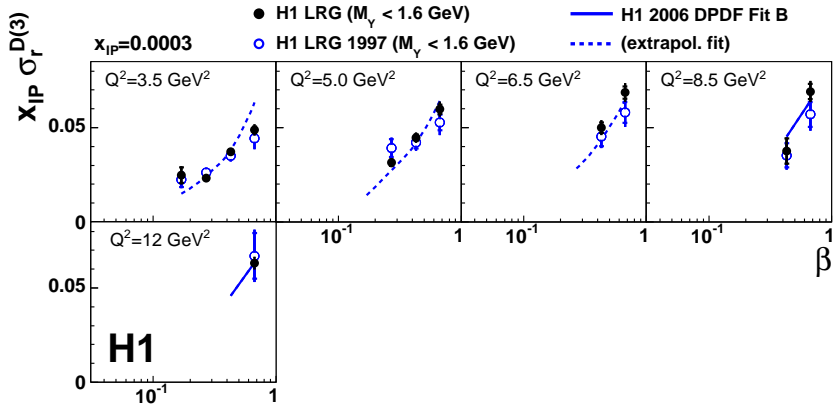}
\end{center}
\vspace*{-2.6cm}
  \caption{The $\beta$ dependence of the reduced diffractive cross section, multiplied by $\xpom$, at a fixed value of  $x_{\pom}= 0.0003$, resulting from the combination of all data samples. 
Previously published H1 measurements~\cite{Aktas:2006hy} are also displayed as open points. 
The inner and outer error bars on the data points represent the statistical and total uncertainties, respectively. Overall normalisation uncertainties of $4\%$ and $6.2\%$ on the combined and previous data, respectively, are not shown.
Predictions from the H1 $2006$ DPDF Fit B~\cite{Aktas:2006hy} are represented by a curve in kinematic regions used to determine the DPDFs and by a dashed line in regions which were excluded from the fit (see section~\ref{sec:models}).}
\label{fig:beta_dep1}
\end{figure}

\begin{figure}[!htbp]
\begin{center}
 \includegraphics[width=.7\textwidth]{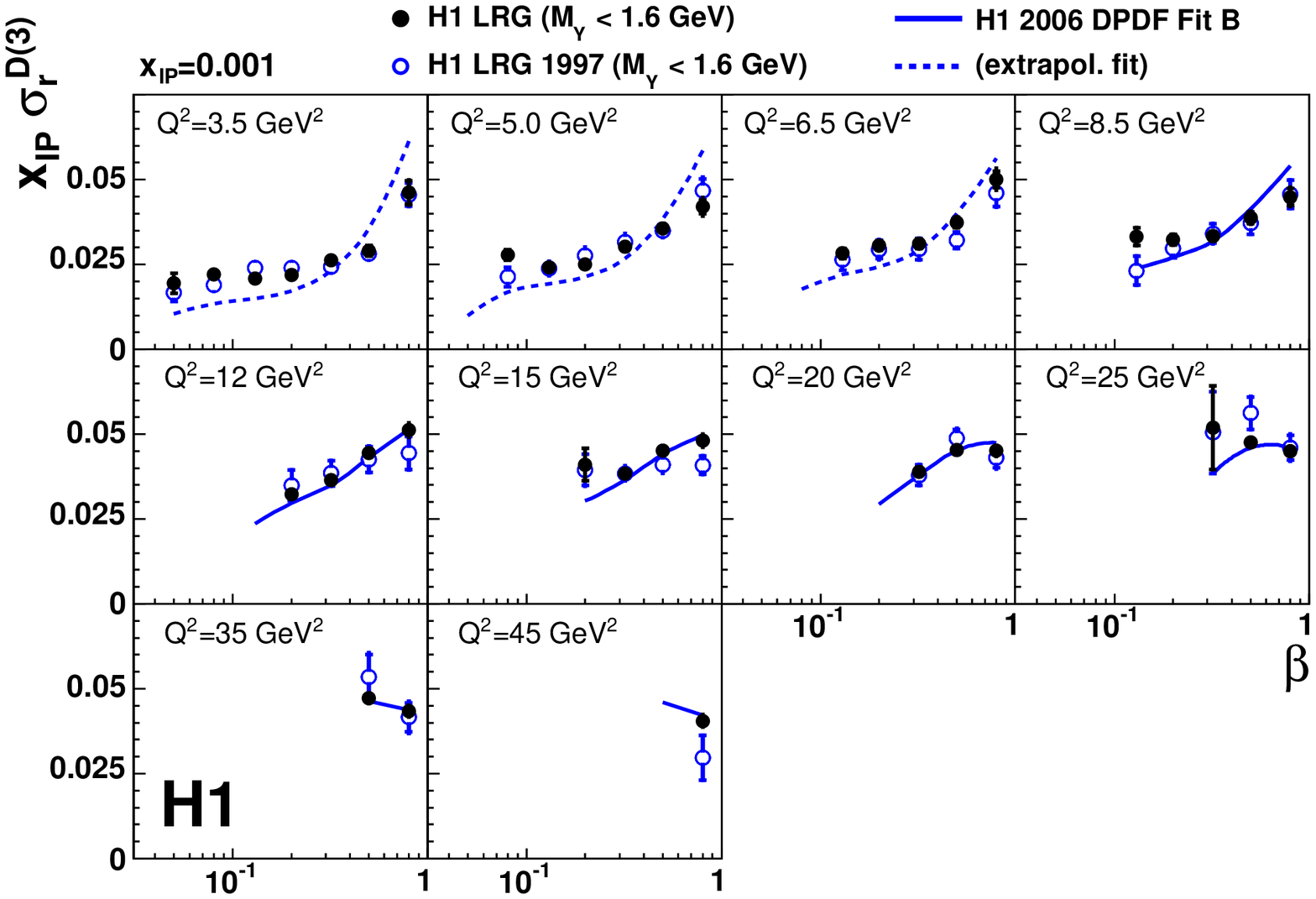}
\end{center}
  \caption{The $\beta$ dependence of the reduced diffractive cross section, multiplied by $\xpom$, at a fixed value of  $x_{\pom}= 0.001$, resulting from the combination of all data samples. Details are explained in the caption of figure~\ref{fig:beta_dep1}.}
\label{fig:beta_dep2}
\end{figure}

\begin{figure}[!htbp]
\begin{center}
 \includegraphics[width=.7\textwidth]{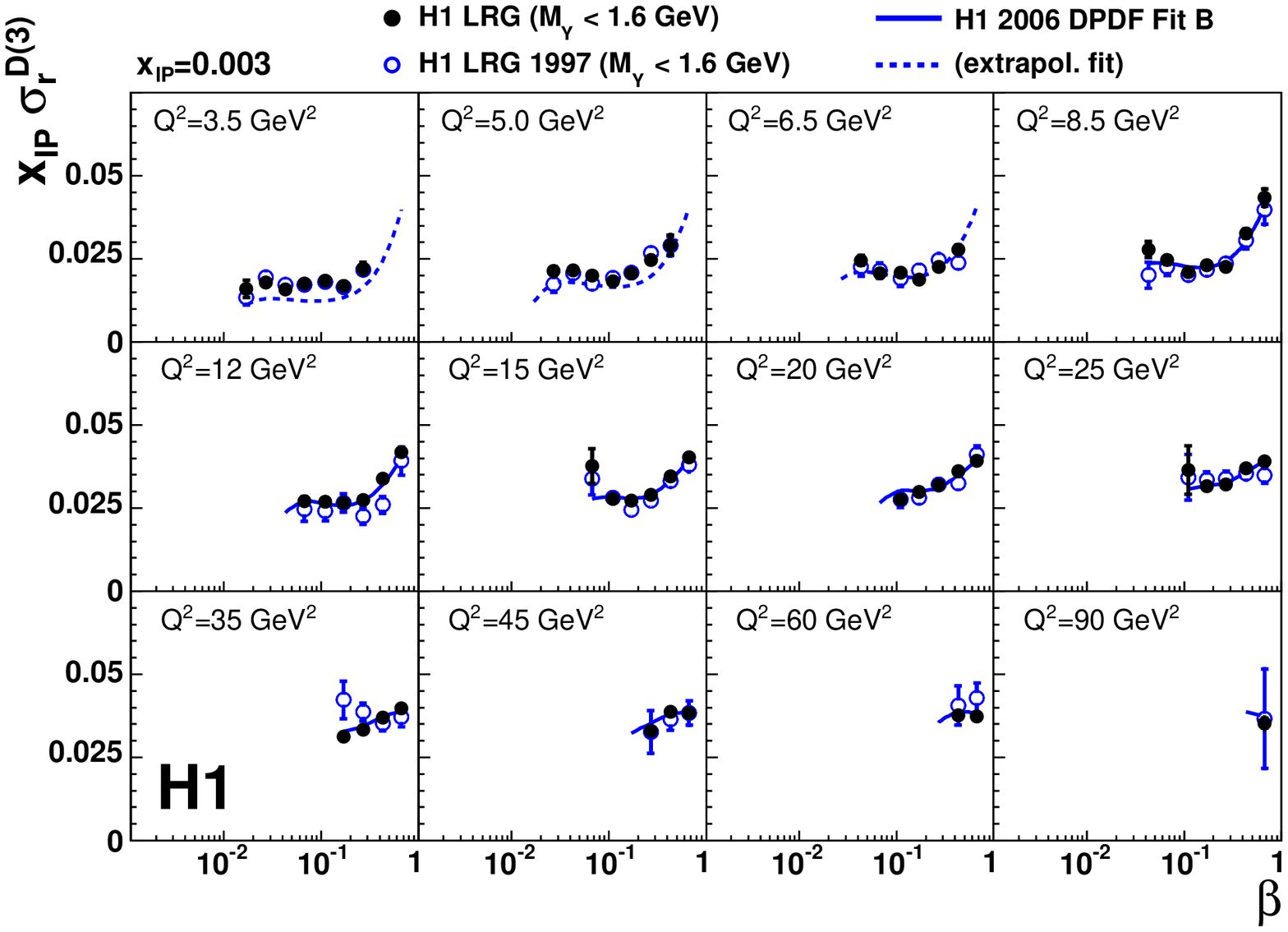}
\end{center}
  \caption{The $\beta$ dependence of the reduced diffractive cross section, multiplied by $\xpom$, at a fixed value of  $x_{\pom}= 0.003$, resulting from the combination of all data samples. Details are explained in the caption of figure~\ref{fig:beta_dep1}.}
\label{fig:beta_dep3}
\end{figure}

\begin{figure}[!htbp]
\begin{center}
 \includegraphics[width=.7\textwidth]{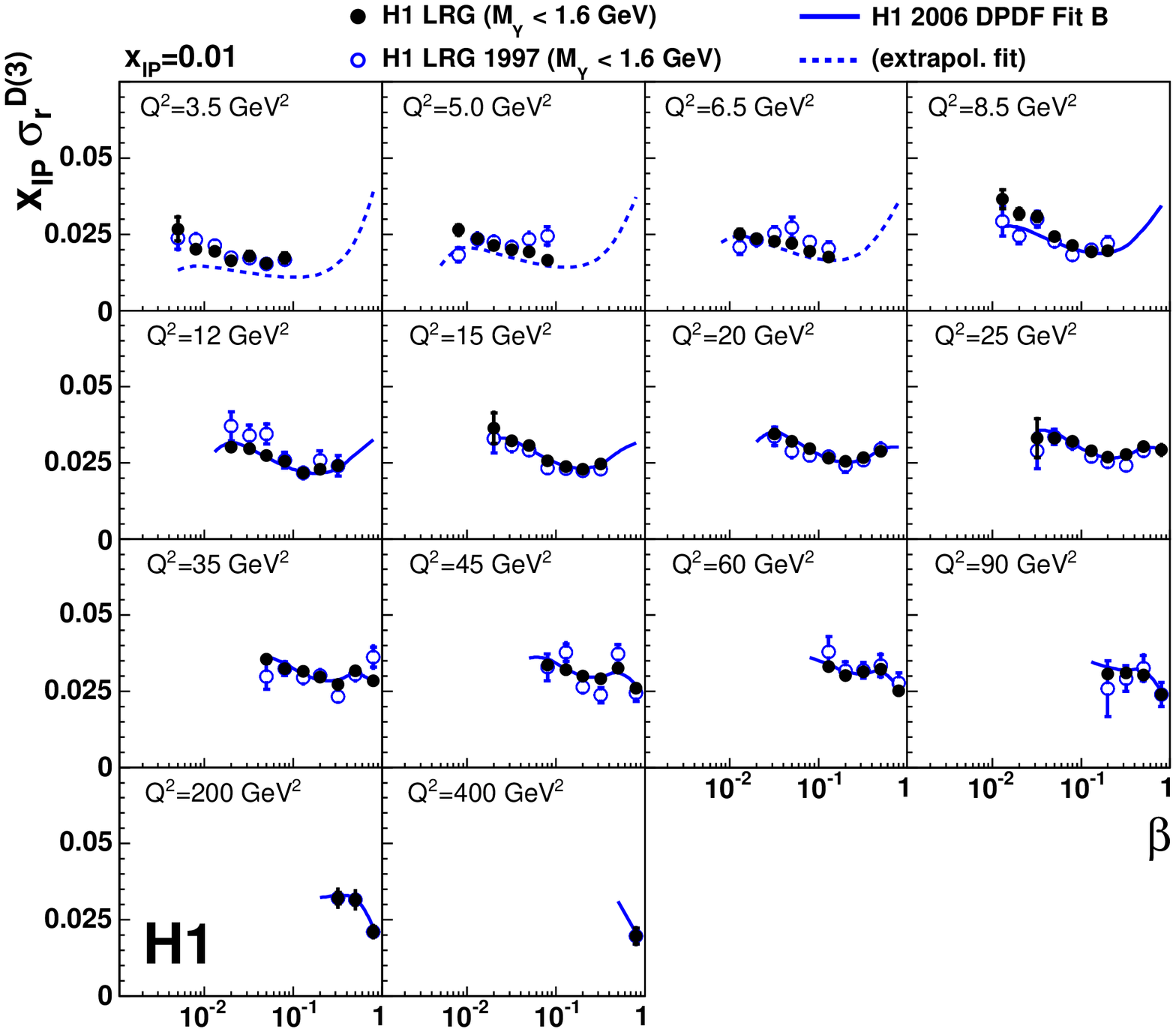}
\end{center}
  \caption{The $\beta$ dependence of the reduced diffractive cross section, multiplied by $\xpom$, at a fixed value of  $x_{\pom}= 0.01$, resulting from the combination of all data samples. Details are explained in the caption of figure~\ref{fig:beta_dep1}.}
\label{fig:beta_dep4}
\end{figure}

\begin{figure}[!htbp]
\begin{center}
 \includegraphics[width=.5\textwidth]{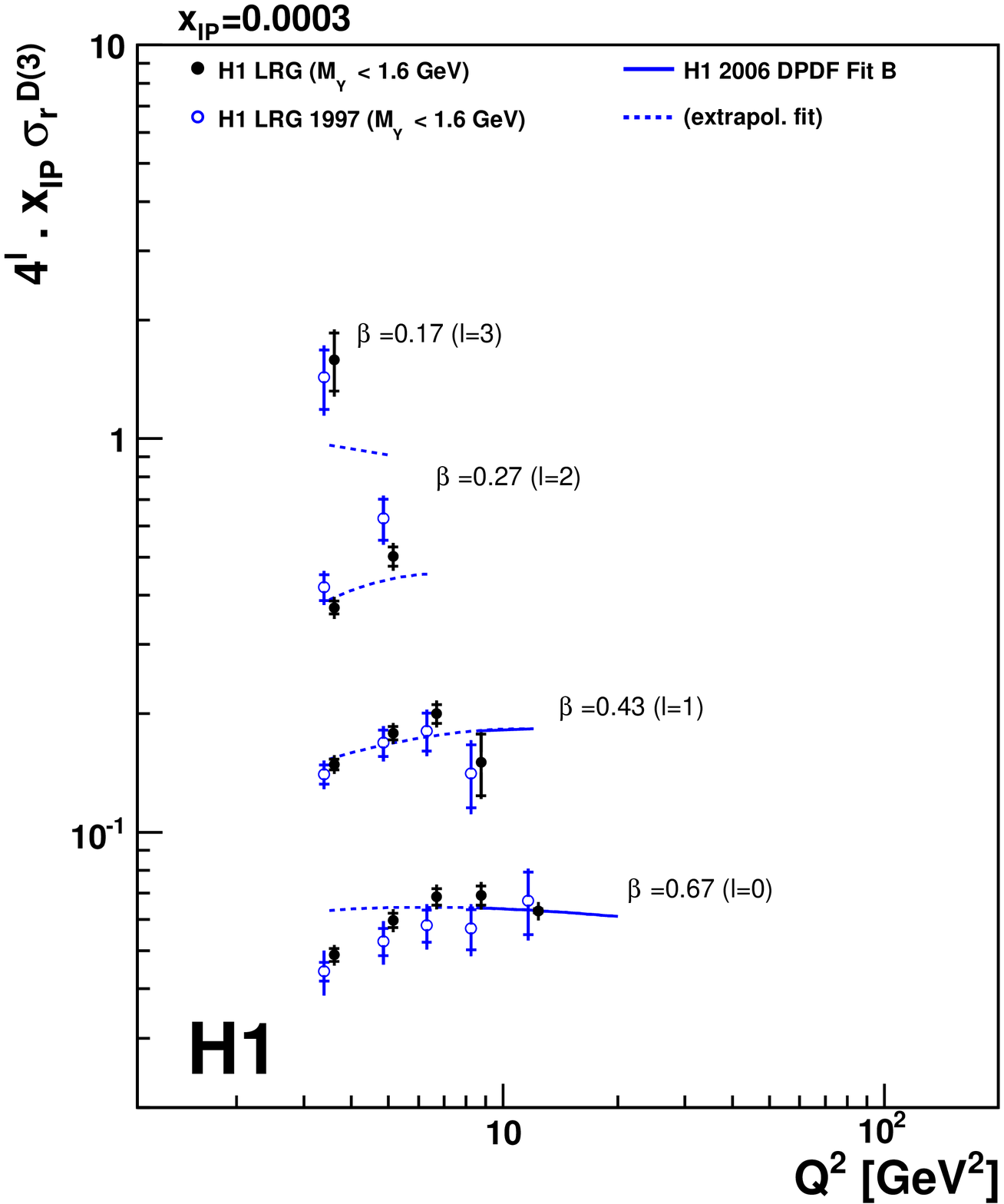}
\put(-7,80){{(a)}}
 \includegraphics[width=.5\textwidth]{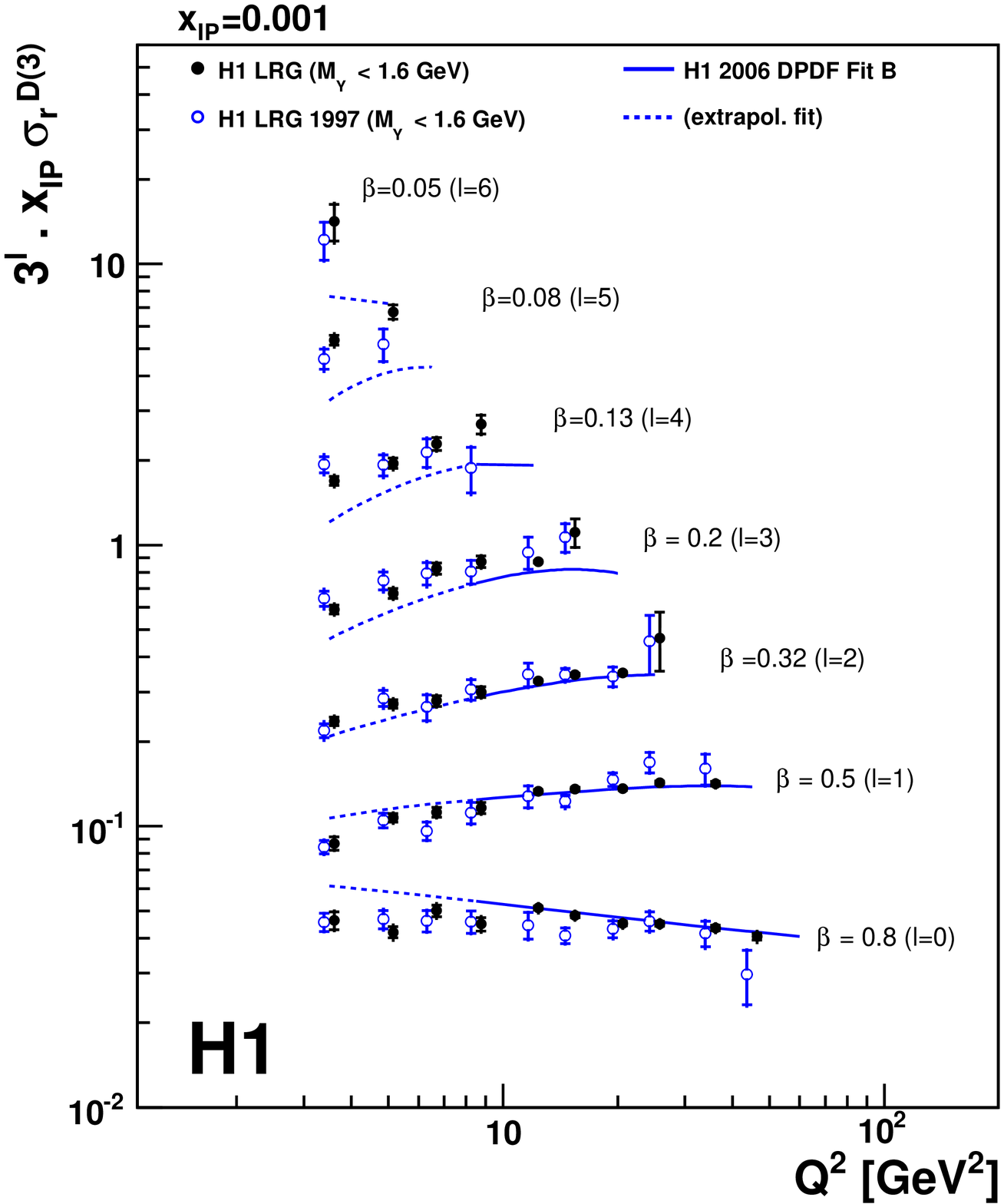}
\put(-7,80){{(b)}}\\
 \includegraphics[width=.5\textwidth]{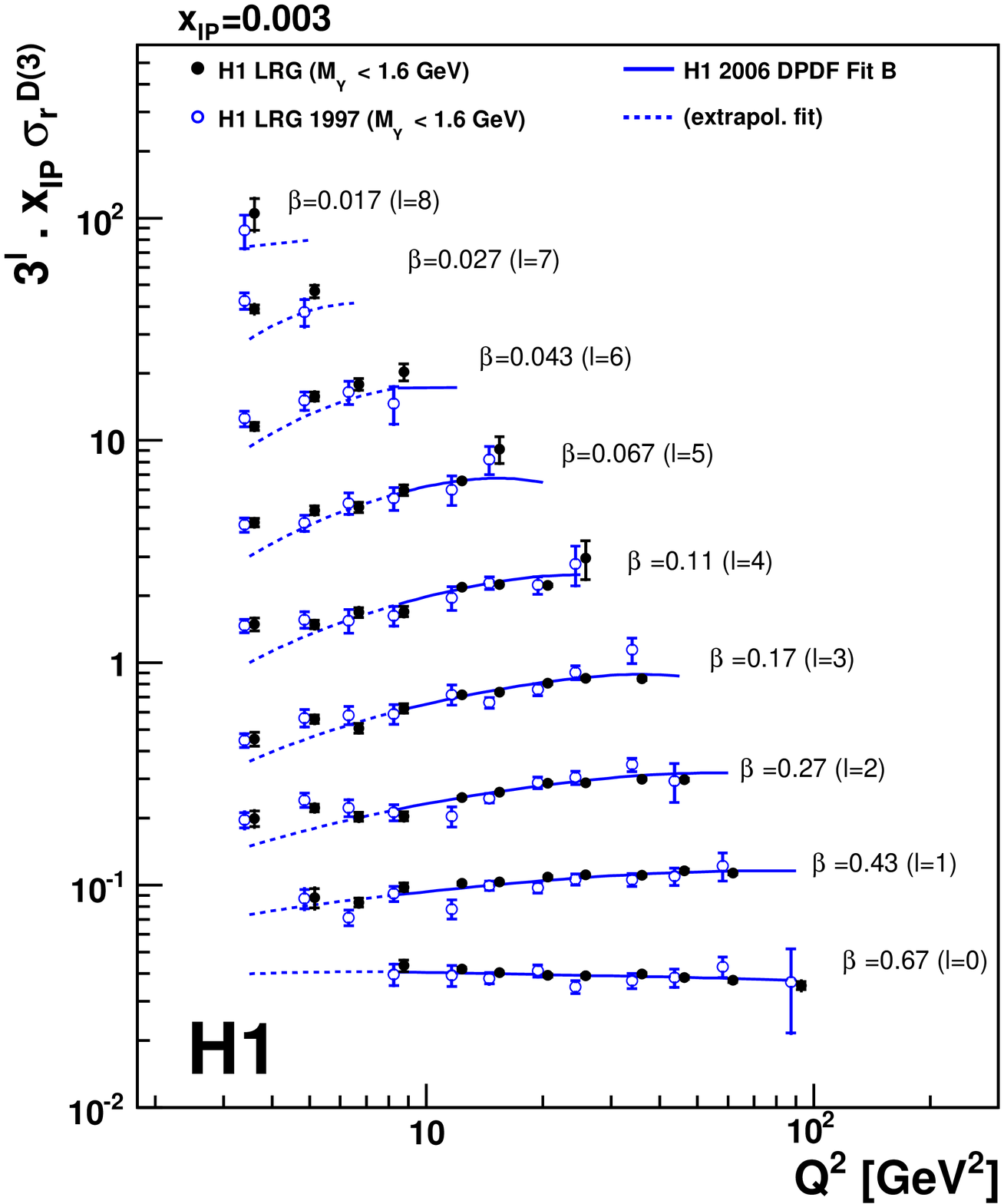}
\put(-7,80){{(c)}}
 \includegraphics[width=.5\textwidth]{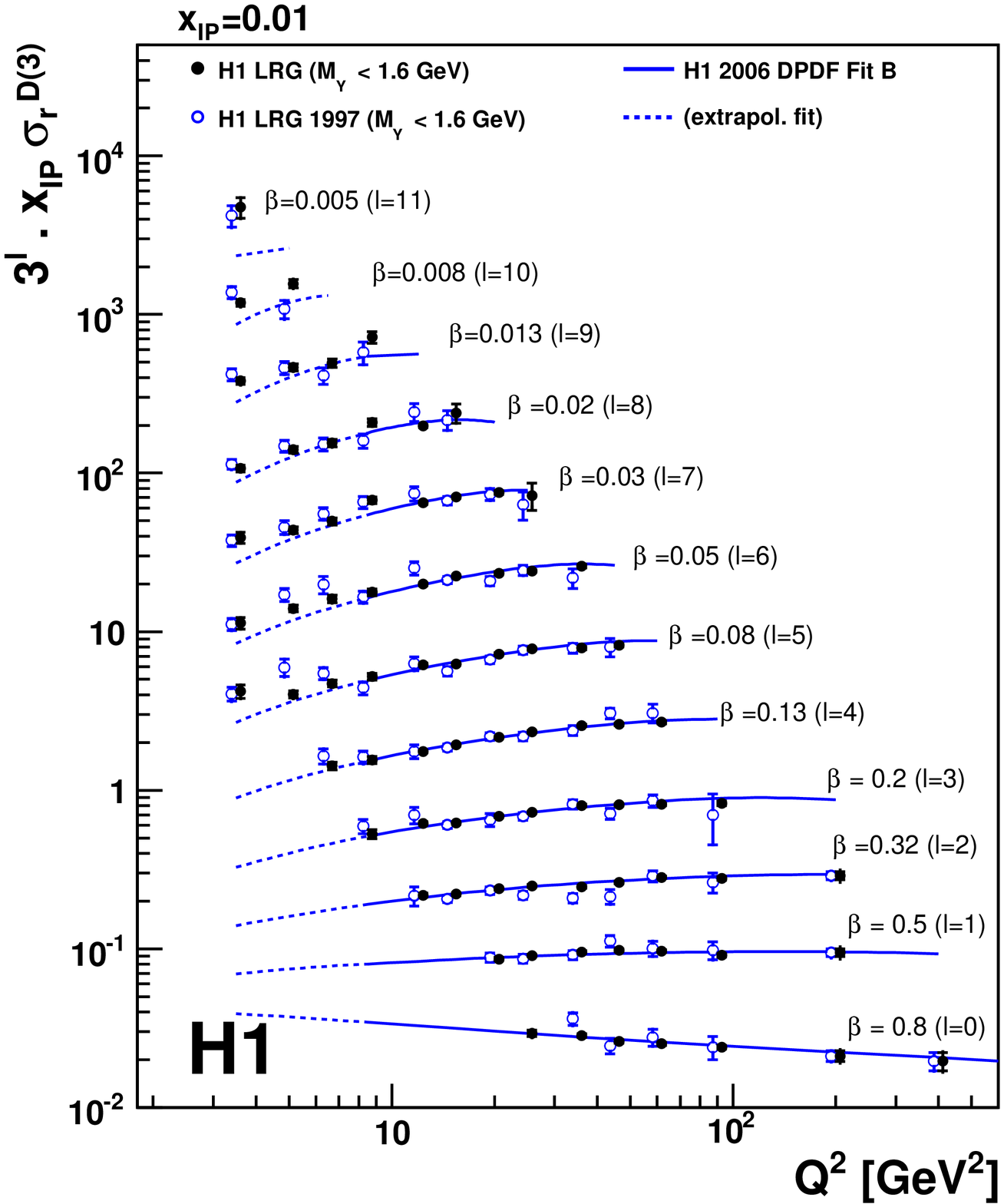}
\put(-7,80){{(d)}}\\
\end{center}
  \caption{The $Q^2$ dependence of the reduced diffractive cross section, multiplied by $\xpom$, at different fixed values of  $x_{\pom}= 0.0003$ (a), $0.001$ (b), $0.003$ (c) and $0.01$ (d), resulting from the combination of all data samples. 
The reduced cross section values are multiplied by a scaling factor, $4^l$ for $\xpom=0.0003$ and $3^l$ for $\xpom=0.003$, $0.001$ and $0.01$, with $l$ values as indicated in parentheses.
Previously published H1 measurements~\cite{Aktas:2006hy} are also displayed as open points. 
The measurements are 
displaced horizontally for better visibility.
More details are explained in the caption of figure~\ref{fig:beta_dep1}.}
\label{fig:q2_dep}
\end{figure}

\begin{figure}[!htbp]
\begin{center}
 \hspace*{0.2cm}\includegraphics[width=.55\textwidth]{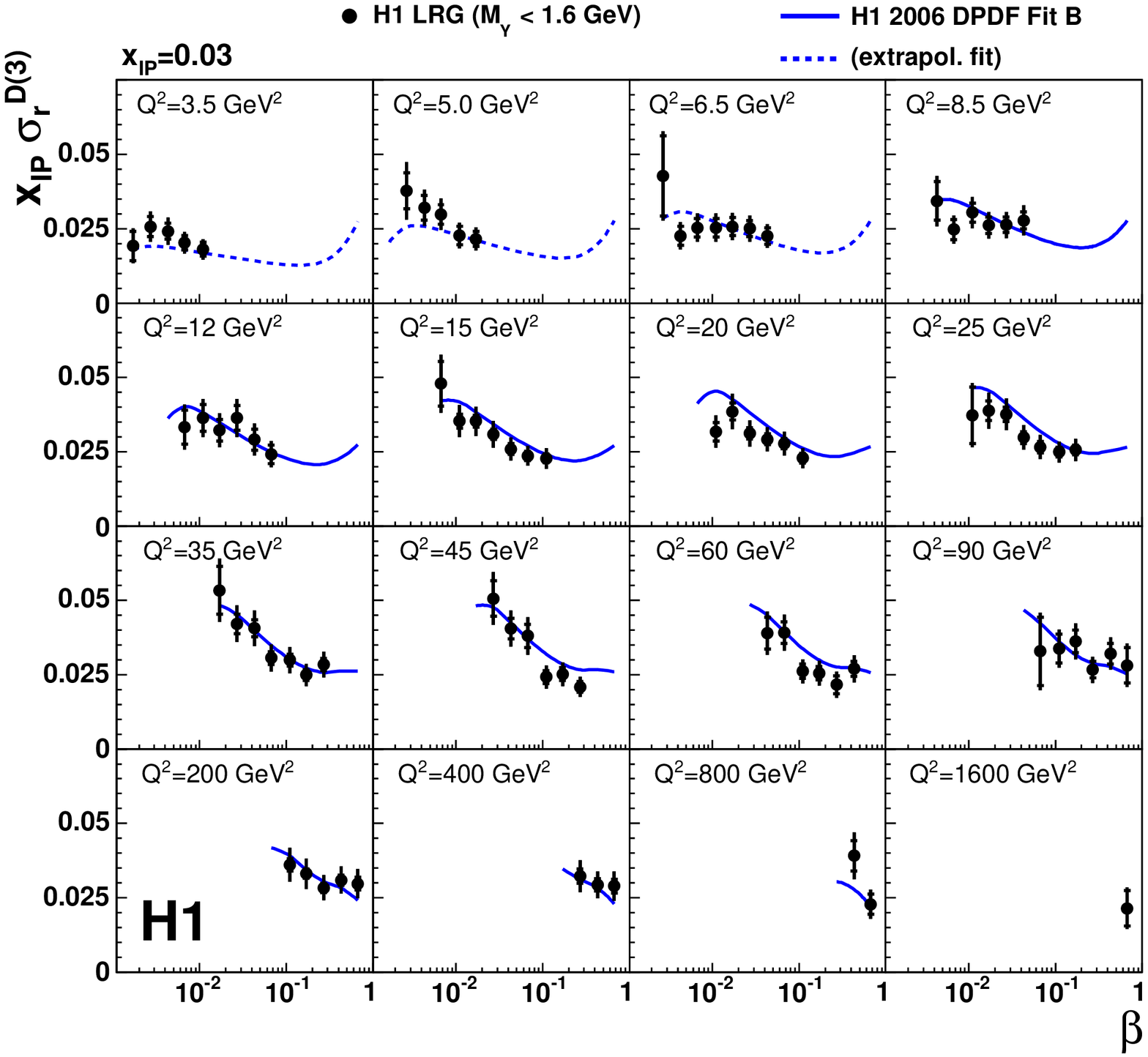}
\put(-10,82){{(a)}}
\end{center}
\begin{center}
 \includegraphics[width=.55\textwidth]{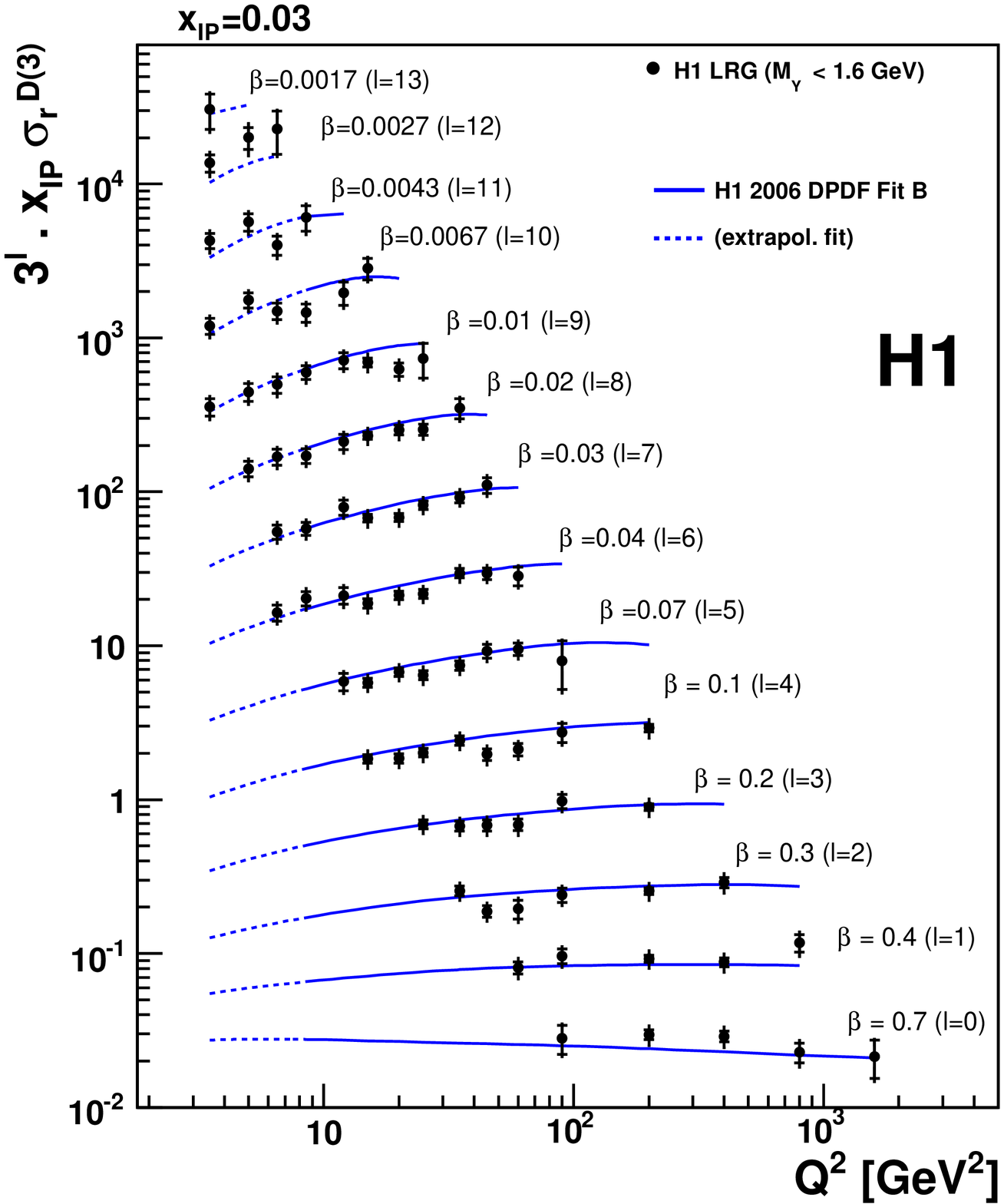}
\put(-7,82){{(b)}}
\end{center}
  \caption{The $\beta$ (a) and $Q^2$ (b) dependences of the reduced diffractive cross section, multiplied by $\xpom$, at a fixed value of  $x_{\pom}= 0.03$, resulting from the combination of all data samples. Details are explained in the caption of figures~\ref{fig:beta_dep1} and~\ref{fig:q2_dep}.}
\label{fig:beta_dep5}
\end{figure}

\begin{figure}[!htbp]
  \begin{center}
 \includegraphics[width=.5\textwidth]{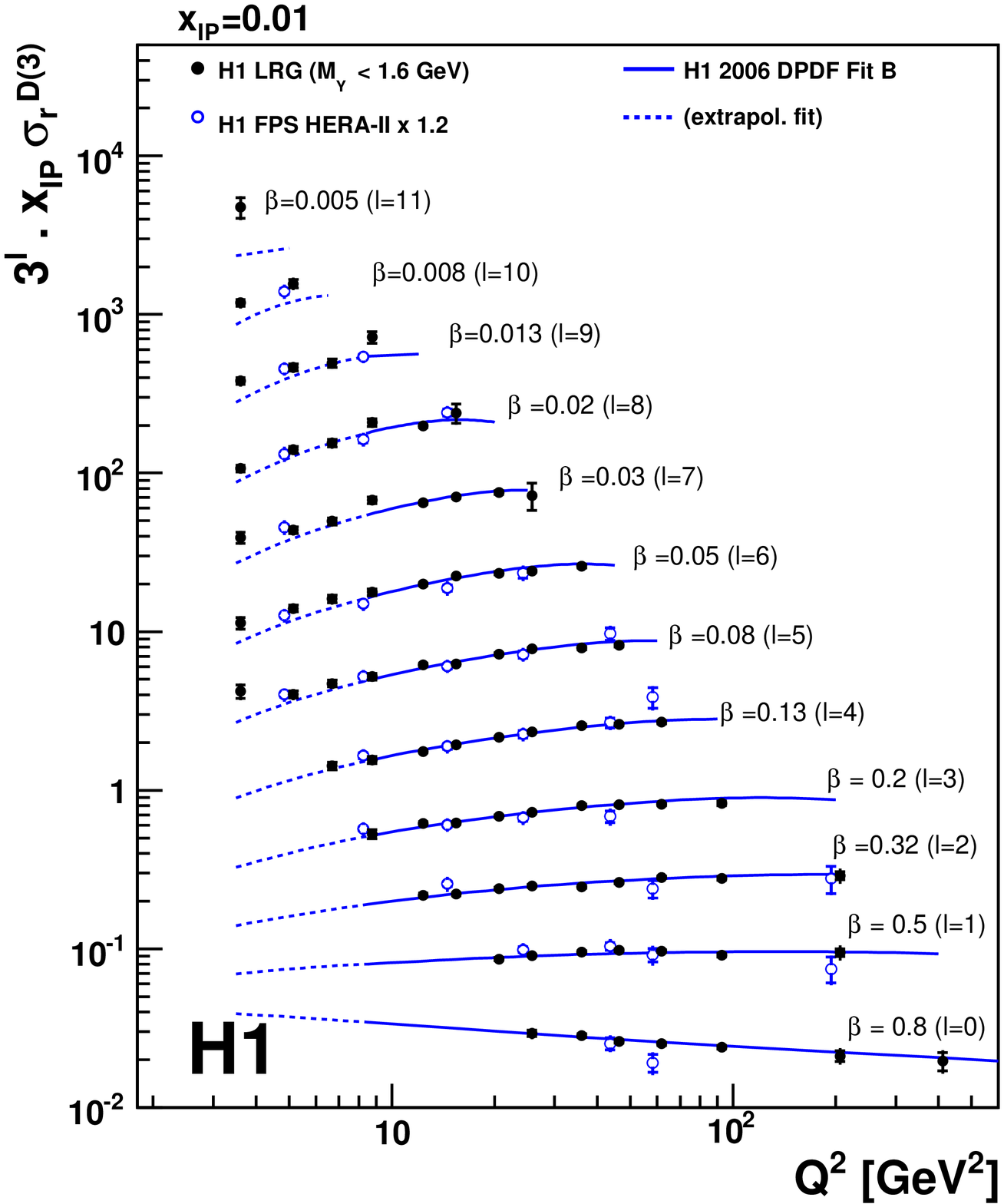}
\put(-7,55){{(a)}}
 \includegraphics[width=.5\textwidth]{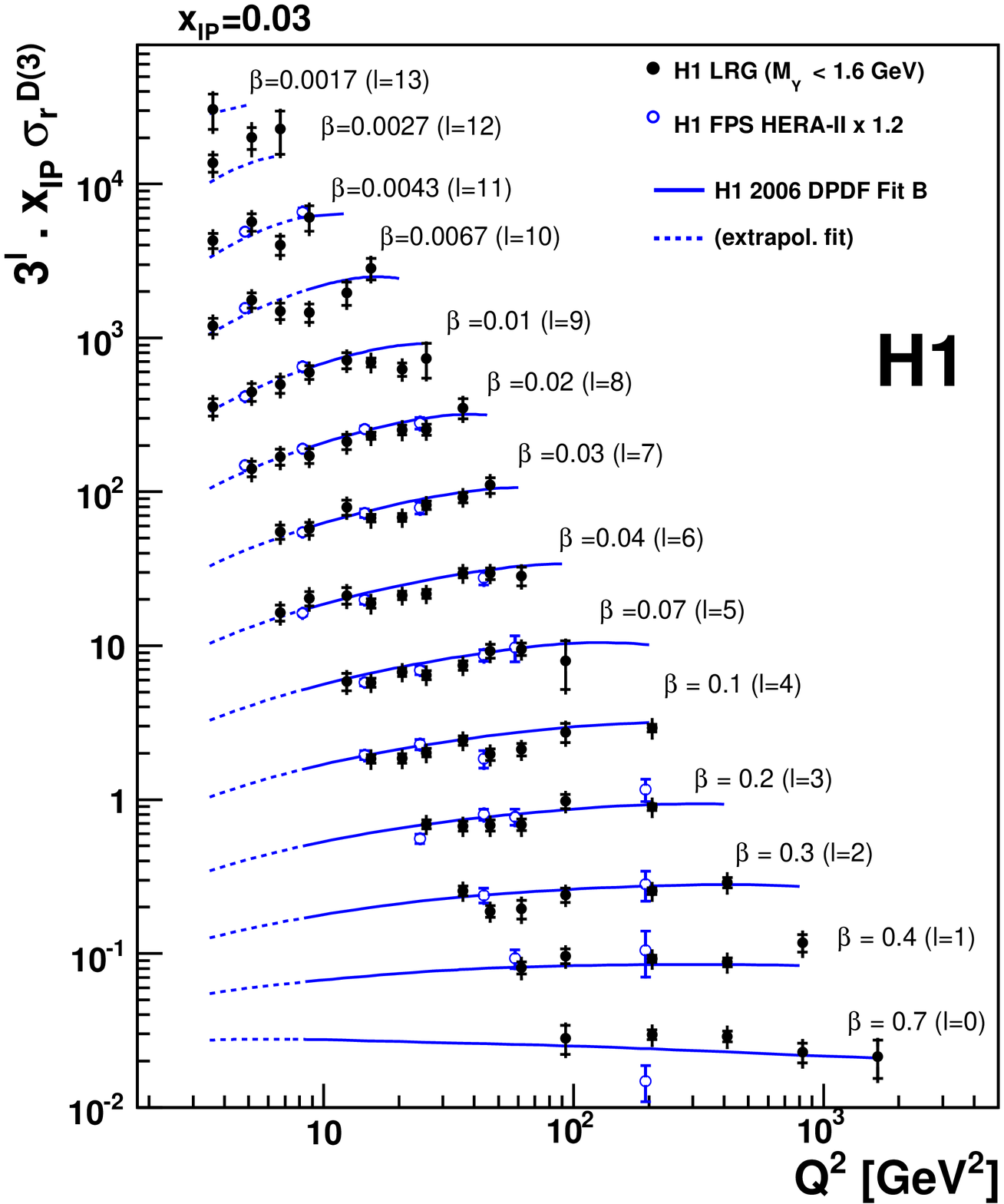}
\put(-7,55){{(b)}}\\
  \end{center}
  \caption{
  The reduced diffractive cross section from combined H1 LRG data,
multiplied by $\xpom$, at two fixed values of
$x_{\pom}$= $0.01$ (a) and $0.03$ (b). 
The reduced cross section values are multiplied by a scaling factor $3^l$, with $l$ values as indicated in parentheses.
The LRG data are compared with the H1 FPS results~\cite{micha} interpolated to the LRG $\beta$, $Q^2$ and $\xpom$ values using a parametrisation of the H1 $2006$ DPDF Fit B~\cite{Aktas:2006hy}. The FPS data are multiplied by a factor $1.2$ (see section~\ref{sec:FPS_ZEUS}). 
The overall normalisation uncertainties of $4\%$ and $6\%$ on the LRG and FPS data, respectively, are not shown.
The measurements are  
displaced horizontally for better visibility.
More details are explained in the caption of figure~\ref{fig:beta_dep1}.
}
\label{fig:FPS}
\end{figure}

\begin{figure}[!htbp]
\begin{center}
 \includegraphics[width=.5\textwidth]{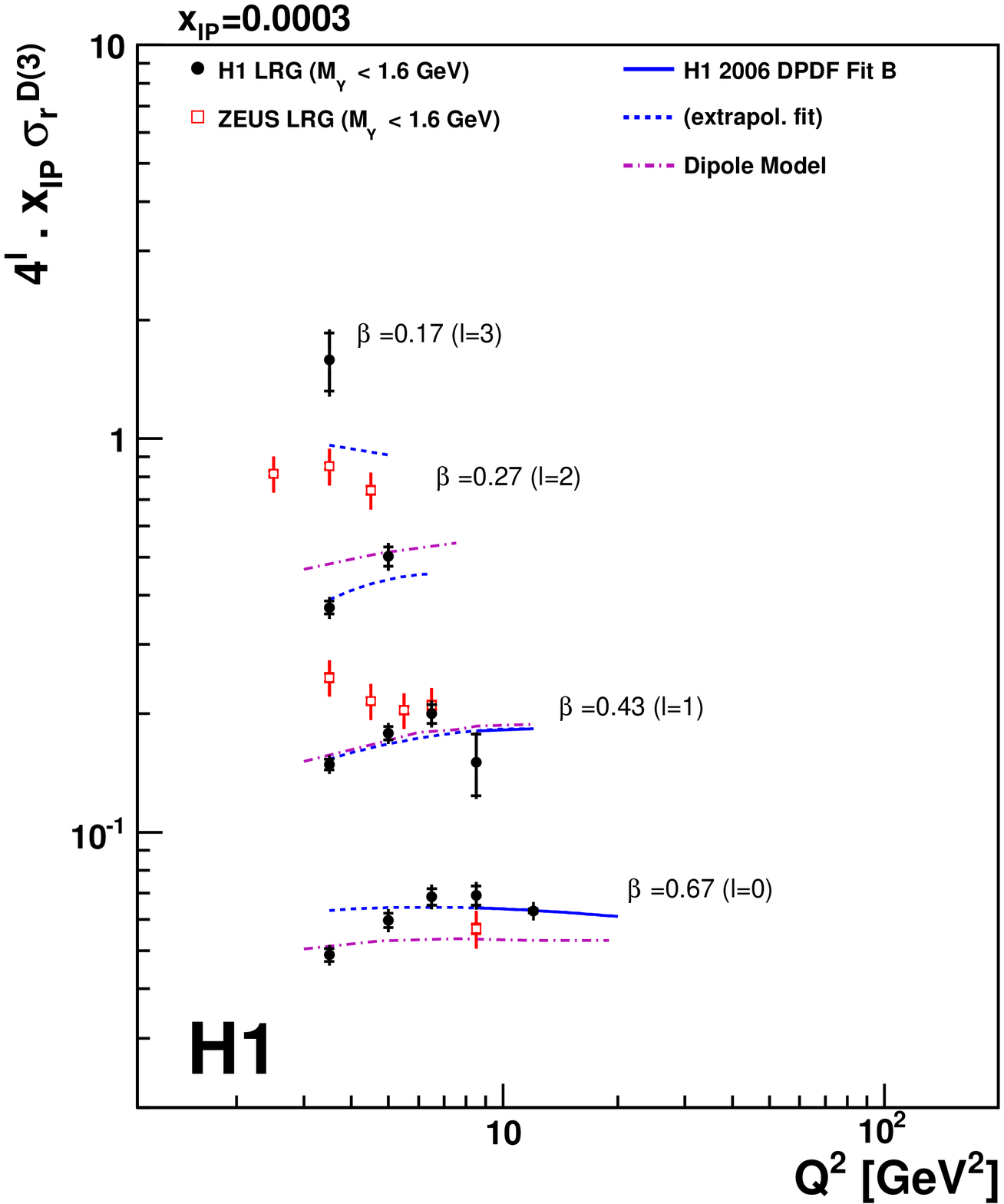}
\put(-7,70){{(a)}}
 \includegraphics[width=.5\textwidth]{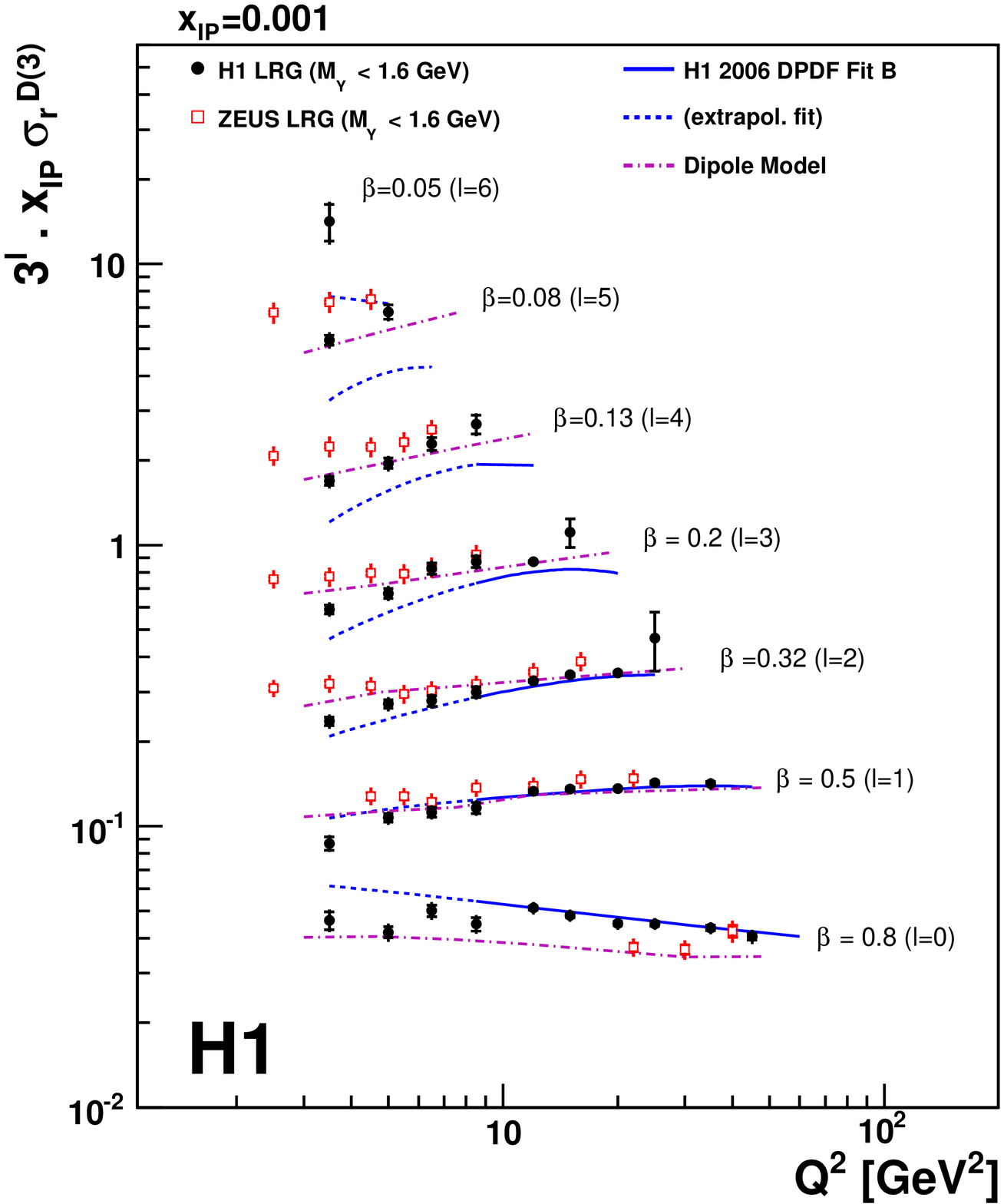}
\put(-7,70){{(b)}}\\
 \includegraphics[width=.5\textwidth]{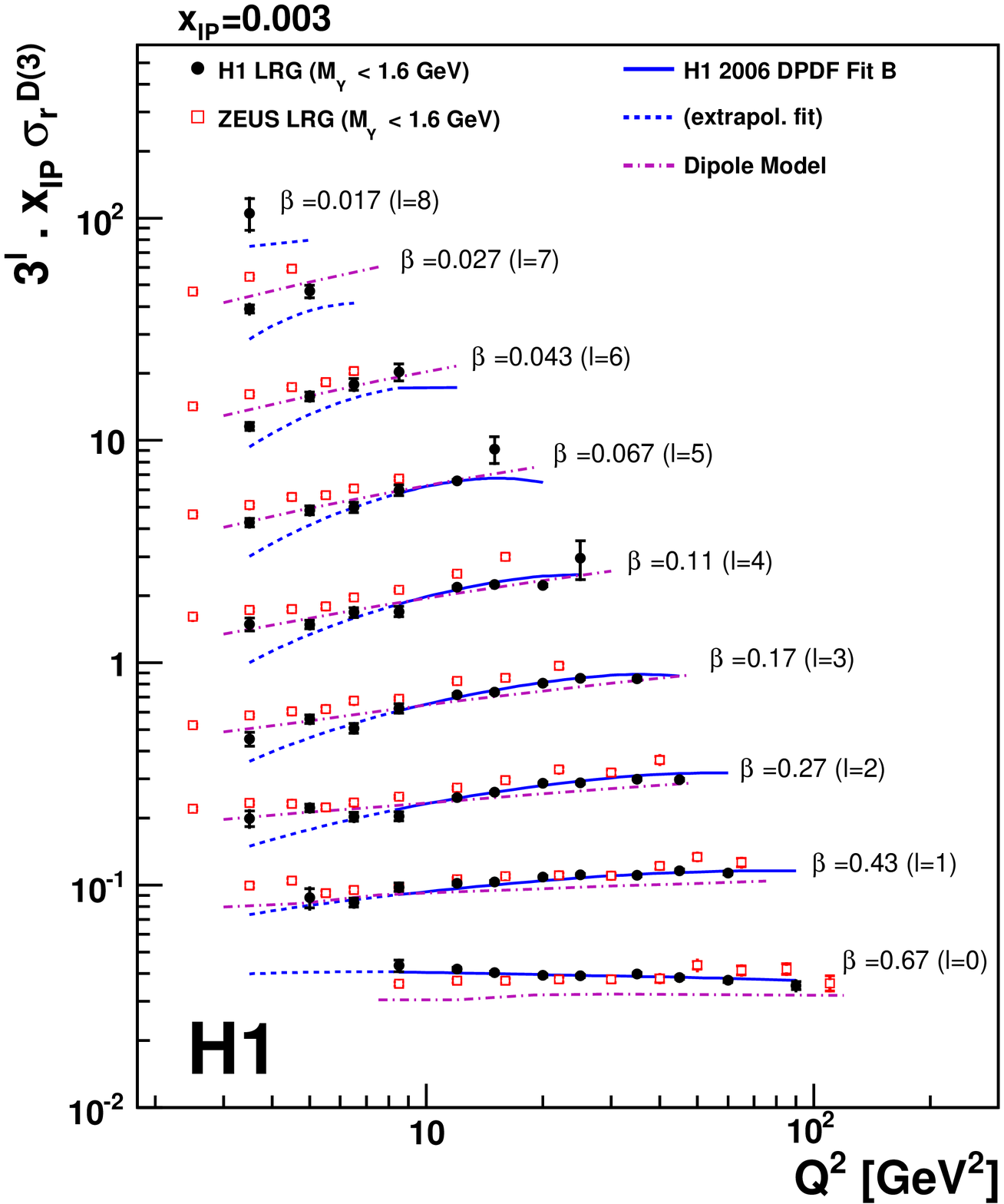}
\put(-7,70){{(c)}}
 \includegraphics[width=.5\textwidth]{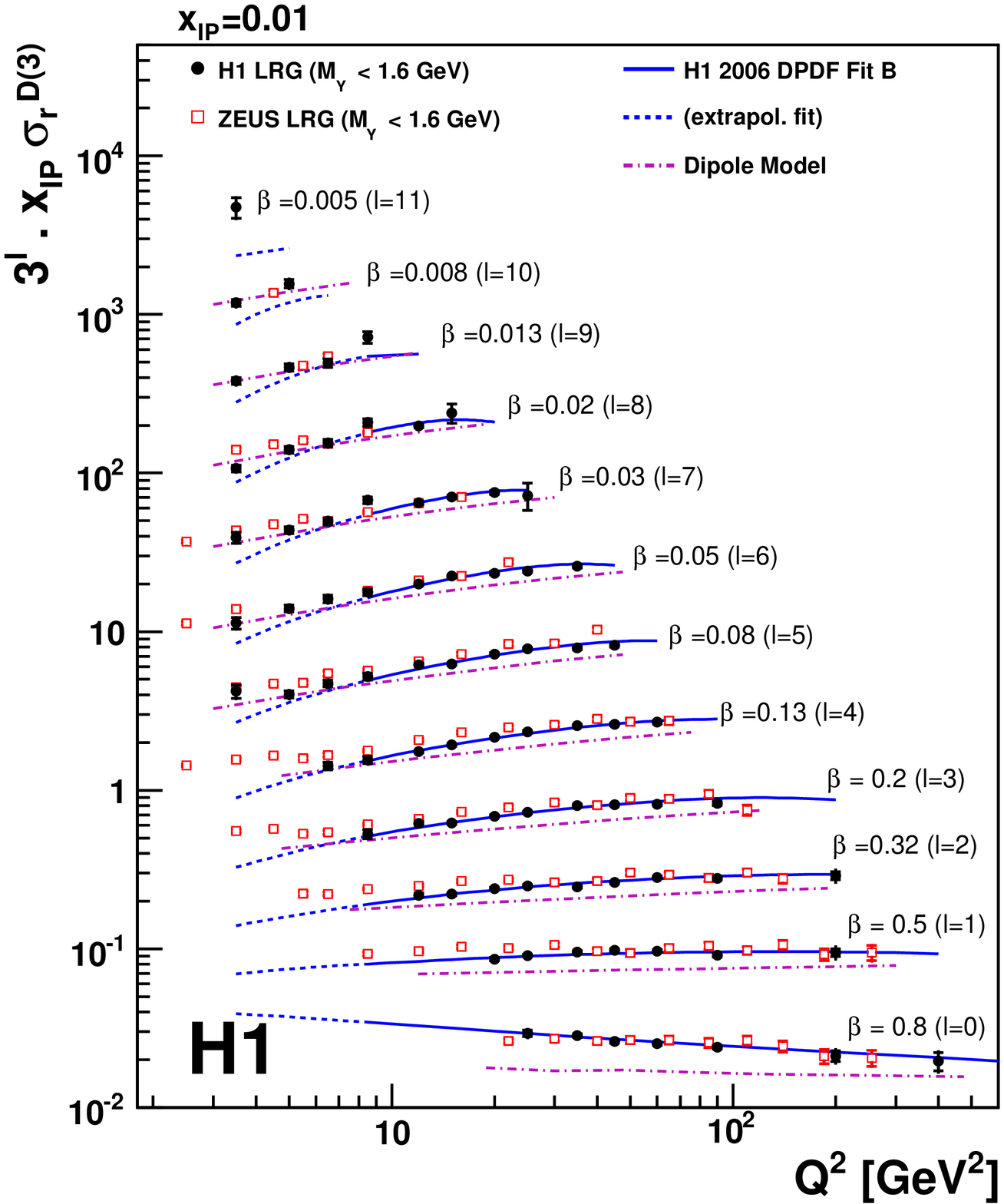}
\put(-7,70){{(d)}}\\
\end{center}
  \caption{The $Q^2$ dependence of the
reduced diffractive cross section from combined H1 data,
multiplied by $\xpom$, at different fixed values of
$x_{\pom}$= $0.0003$ (a), $0.001$ (b), $0.003$ (c) and $0.01$ (d). 
The present data are compared with the results of the ZEUS Collaboration~\cite{Chekanov:2008fh}, corrected to $M_Y < 1.6$~GeV (see text). 
The $8\%$ overall uncertainty on this correction for ZEUS data is not shown.
The overall normalisation uncertainties of $4\%$ and $2.25\%$ for the H1 and ZEUS data, respectively, are also not shown.
Predictions from the H1 $2006$ DPDF Fit B~\cite{Aktas:2006hy} and dipole model~\cite{Marquet:2007nf} are displayed.
More details are explained in the captions of figures~\ref{fig:beta_dep1} and~\ref{fig:q2_dep}.}
\label{fig:ZEUS}
\end{figure}

Figures~\ref{fig:beta_dep1} to~\ref{fig:ZEUS} show the measurements compared to predictions 
based on the H1 $2006$ DPDF Fit B.
The DPDF fit assuming proton vertex factorisation  used in the previous H1 analysis~\cite{Aktas:2006hy} became unstable when data points with $Q^2 < 8.5$~GeV$^2$ were included. 
Therefore, only an extrapolation of the DPDFs predictions to this kinematic domain is indicated as dashed lines in these figures.
In figure~\ref{fig:ZEUS} the data are compared also with predictions of the  dipole model~\cite{Marquet:2007nf}.
As the dipole model predictions correspond to the process $ep \rightarrow eXp$,  they are rescaled by a factor of $1.20$ according to equation~(\ref{eq:LRG_over_FPS}).
Both approaches give a good overall description of the measurements.
In the low $Q^2$ range, for $Q^2 < 8.5$~GeV$^2$, the dipole model, which includes saturation effects, seems to better describe the data, whereas for larger $\beta$ and for $\xpom = 0.01$ it tends to underestimate the measured cross section.

\subsection{Ratio to Inclusive DIS}

\begin{figure}[p]
  \center
  \includegraphics[width=.99\textwidth]{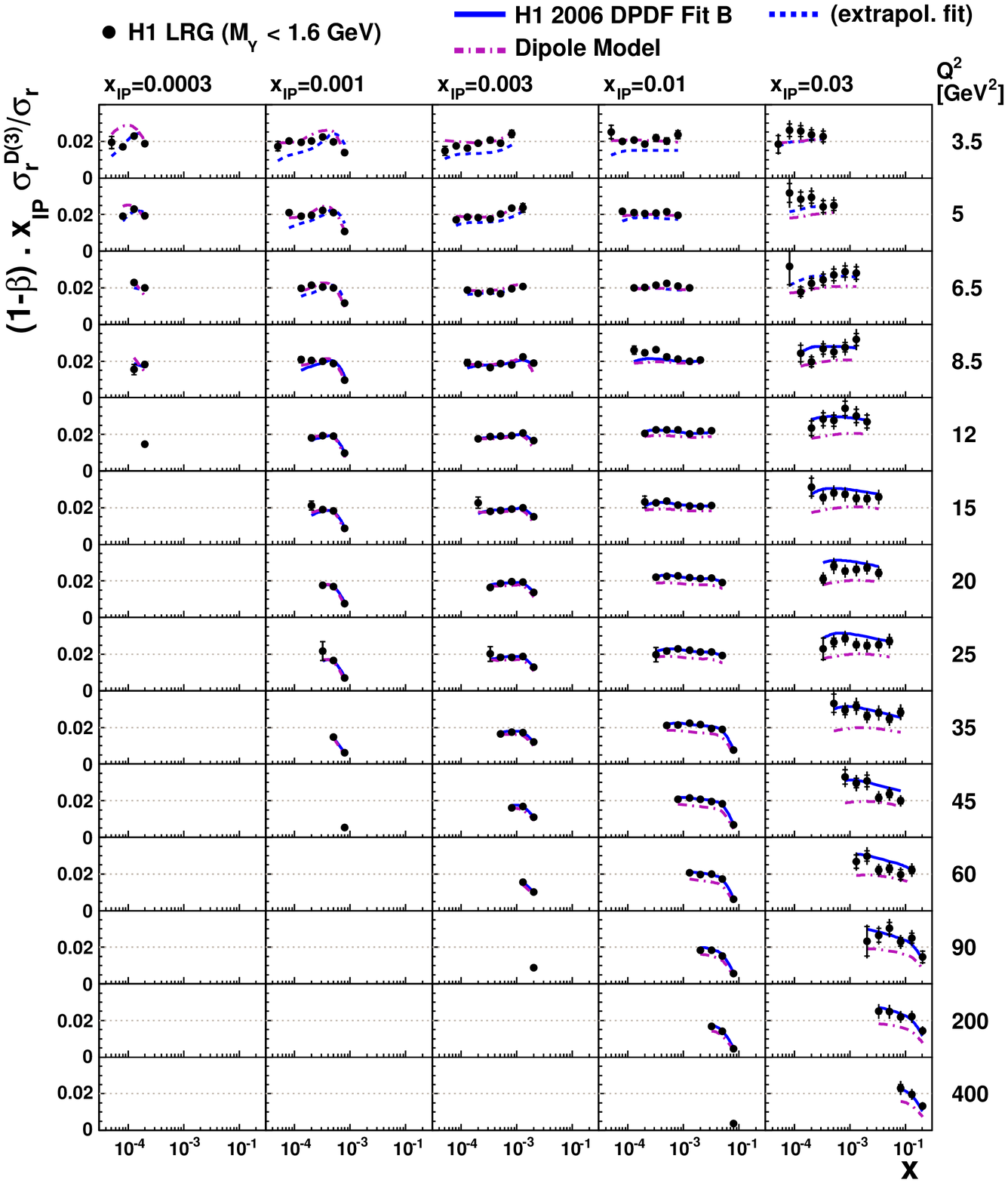}
  \caption{The ratio of the diffractive to the inclusive reduced cross section, multiplied by $(1-\beta)\xpom$. The inner and outer error bars on the data points represent the statistical and total uncertainties, respectively. The overall normalisation uncertainty of $4\%$ is not shown. 
  The curves are explained in the captions of figures~\ref{fig:beta_dep1} and~\ref{fig:ZEUS}.}
  \label{fig:F2DoverF2}
\end{figure}

In analogy to hadronic scattering, the diffractive and the total cross sections can be related via the generalisation of the optical theorem to virtual photon scattering~\cite{optic_th}. 
Many models of low $x$ DIS~\cite{lowx1,lowx2,lowx3,lowx4,lowx5,lowx6} assume links between these quantities.
Comparing the $Q^2$ and $x$ dynamics of the diffractive with the inclusive cross section is therefore a powerful means of comparing the properties of the DPDFs with their inclusive counterparts and of testing models.
The evolution of the diffractive reduced cross section with $Q^2$ can be compared with that of the inclusive DIS reduced cross section $\sigma_r$ by forming the ratio 
\begin{equation}
\frac{\sigma^{D(3)}_r(x_\pom,x,Q^2)}{\sigma_r(x,Q^2)}\, . \, (1-\beta) \, x_\pom \, ,
\label{eq:diff_over_incl}
\end{equation} 
\noindent at fixed $Q^2$, $\beta = x/x_\pom$ and $x_\pom$. A parametrisation of $\sigma_r$ from~\cite{Aaron:2009kv} is used.
This quantity is equivalent to the ratio of diffractive to 
$\gamma^*p$~cross sections, 
\begin{equation}
\frac{M_X^2 \,\, \frac{\displaystyle {\rm d} \sigma_r^{D(3)}(M_X, W, Q^2)}{\displaystyle {\rm d} M_X}}{ \sigma_{incl.}^{\gamma^* p}(W,Q^2)} \, ,
\end{equation}
\noindent studied in~\cite{zeuslps,ZEUS:97,Chekanov:2008cw} as a function of $W$ and $Q^2$ in ranges of $M_X$.
Assuming proton vertex factorization in the DPDF approach, this ratio is expected to be independent of $Q^2$ and  depends only weakly on $\beta$ and $x \simeq Q^2 / W^2$ for sufficiently large $M_X$. 
A remaining weak $x$ dependence of the ratio may arise due to deviations from unity of the intercept of the Pomeron trajectory, which are studied in the next section.
The ratio~(\ref{eq:diff_over_incl}) is shown in figure~\ref{fig:F2DoverF2} as a function of $x$ at fixed $x_\pom$ and $Q^2$ values.
The ratio of the diffractive to the inclusive cross section is found to be approximately constant with $x$ at fixed $Q^2$ and $x_\pom$ except towards larger $x$ values which correspond to large $\beta$ values. 
This indicates that the ratio of quark to gluon distributions is similar in the diffractive and inclusive process when considered at the same low $x$ value.
The ratio is also larger at high values of $x_\pom$, $\xpom = 0.03$, where the sub-leading exchange contribution of the diffractive cross section is not negligible, but it remains approximately constant with $x$.
These observations are in agreement with previous similar studies~\cite{micha}.
The general behaviour of the ratio, and especially its decrease towards larger $x$, is reproduced by both the DPDF~\cite{Aktas:2006hy} and dipole model~\cite{Marquet:2007nf}  predictions.

\subsection{Extraction of the Pomeron Trajectory}

The diffractive structure function $F_2^{D(3)}$ is obtained from the reduced cross section by correcting for the small $F_L^{D(3)}$ contribution using the predictions of the H1 $2006$ DPDF Fit B, which is in reasonable agreement with the recent direct measurement of $F_L^{D(3)}$~\cite{FLDpaper}. 
The diffractive structure function can be investigated in the framework of Regge phenomenology and is usually expressed as a sum of two factorised contributions corresponding to Pomeron and secondary Reggeon trajectories
\begin{eqnarray}
F_2^{D(3)}(Q^2,\beta,x_{\PO})=
f_{\PO / p} (x_{\PO}) \;  F_2^{\PO} (Q^2,\beta)
+n_{\RO}  \; f_{\RO / p} (x_{\PO}) \; F_2^{\RO} (Q^2,\beta) \ .
\label{reggeform}
\end{eqnarray}
\noindent In this parametrisation, $F_2^{\PO}$ can be interpreted as the Pomeron structure function  and $F_2^{\RO}$ as an effective Reggeon structure function.
The global normalisation of this last contribution is denoted $n_{\RO}$.
The Pomeron and Reggeon fluxes are assumed to follow a Regge behaviour with  
linear trajectories $\alpha_{\PO,\RO}(t)=\alpha_{\PO,\RO}(0)+\alpha^{'}_{\PO,\RO} t$, such that
\begin{equation}
f_{{\PO} / p,{\RO} / p} (x_{\PO})= \int^{t_{min}}_{t_{cut}} 
\frac{e^{B_{{\PO},{\RO}}t}}
{x_{\PO}^{2 \alpha_{{\PO},{\RO}}(t) -1}} {\rm d} t .
\label{flux}
\end{equation}
\noindent In this formula, $|t_{min}|$ is the minimum kinematically allowed value of $|t|$ and $t_{cut}=-1$ GeV$^2$ is the limit of the measurement.

\begin{figure}[htbp]
  \begin{center}
    \includegraphics[width=.7\textwidth]{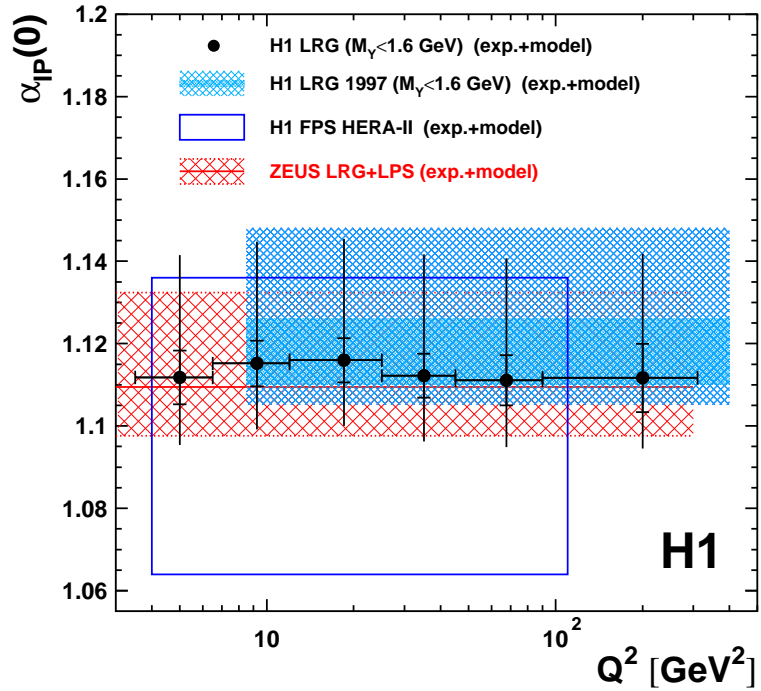}
  \end{center}
  \caption{ Pomeron intercept values
  obtained from Regge fits in different $Q^2$ bins, as defined in the text (dots).
The inner error bars represent the statistical and systematic errors added in quadrature and the outer error bars include model uncertainties in addition (see text for details).
  Previous determinations of the Pomeron intercept~\cite{Aktas:2006hy,Aktas:2006hx,micha,Chekanov:2008fh}
  are also displayed for comparison. 
  For these previous results the bands or boxes
  represent the combination of experimental and model uncertainties,
  always dominated by the model error.}
\label{alphapom}
\end{figure}

In equation~(\ref{reggeform}), the values of $F_2^{\PO}$ are treated as  free parameters at each $\beta$ and $Q^2$ point, together with the Pomeron intercept $\alpha_{\PO}(0)$ and the normalisation $n_{\RO}$ of the sub-leading exchange.
The values of the other parameters are fixed in the fit.
The parameters $\alpha^{'}_{\PO}=0.04^{+0.08}_{-0.06}$~GeV$^{-2}$ and $B_{\PO}=5.7^{+0.8}_{-0.9}$ GeV$^{-2}$ are taken from the last H1 FPS publication~\cite{micha}.
The intercept of the sub-leading exchange $\alpha_{\RO}(0)=$~$0.5 \pm 0.1$ is taken from~\cite{h1f2d94}.
The parameters $\alpha^{'}_{\RO}=0.30^{+0.6}_{-0.3}$~GeV$^{-2}$ and $B_{\RO}=1.6^{-1.6}_{+0.4}$~GeV$^{-2}$ are obtained from a parametrisation of previously published H1 FPS data~\cite{Aktas:2006hx}.
Since the sub-leading exchange is poorly constrained by the data, values of $F_2^{\RO} (Q^2,\beta)$ are taken from a parametrisation of the pion structure function~\cite{GRVpion}, with a single free normalisation $n_{\RO}$. 
Choosing a different parametrisation for the pion structure function~\cite{owens} does not affect the results significantly.
In previous publications~\cite{h1f2d94,Aktas:2006hx,micha,Chekanov:2008fh}, it has already been shown that fits of this form provide a good description of the data. 
This supports the proton vertex factorisation hypothesis whereby the $x_{\pom}$ and $t$ dependences are decoupled from the $Q^2$ and $\beta$ dependences for each of the Pomeron and sub-leading contributions.
This global conclusion can be refined using the advantage of the improved statistical precision of the present analysis.
In the following, the full range in $Q^2$ is divided into six intervals:
$Q^2 \le 6.5$ GeV$^2$, $6.5<Q^2 \le 12$ GeV$^2$,
 $12<Q^2 \le 25$ GeV$^2$,
 $25<Q^2 \le 45$ GeV$^2$,
 $45<Q^2 \le 90$ GeV$^2$
and $Q^2 > 90$ GeV$^2$.
For each interval $i$, a free Pomeron intercept $\alpha_{\PO}(0)[Q_i^2]$ is introduced.
Thus the factorisation assumption can be tested differentially in $Q^2$ by allowing for a $Q^2$ dependence of the Pomeron intercept in the fit procedure.
In the minimisation procedure the error of each data points is obtained by adding in quadrature the statistical and uncorrelated systematic uncertainties.
The effect of correlated uncertainties is taken into account by repeating the fit multiple times with each correlated systematic error shifted by one standard deviation.
The kinematic domain of the fit procedure is defined as $M_X>2$ GeV and $\beta<0.8$, in order to avoid resonances and potential higher-twist effects. 
This leads to $175$ diffractive structure function values.
The fit provides a good description of the data ($\chi^2=201$). The results on the Pomeron intercept are presented in figure~\ref{alphapom}.
No significant $Q^2$ dependence of the Pomeron intercept is observed, which supports the proton vertex factorisation hypothesis.
The average value is found to be
\begin{eqnarray}
\alpha_{\PO}(0)= 1.113 \ \pm 0.002 \ \mathrm{(exp.)}  \ 
^{+ 0.029}_{- 0.015} \ \mathrm{(model)} \ , 
\label{alpom:answer}
\end{eqnarray}
\noindent where the first error is the full experimental uncertainty and the second error expresses the model 
dependent uncertainty arising dominantly from the variation of $\alpha_{\PO}^\prime$, which is strongly positively correlated with $\alpha_{\PO}(0)$. 
As diffractive structure function values are determined with an assumption on $F_L^{D(3)}$, the influence of neglecting the $F_L^{D(3)}$  contribution is also included in the model dependent uncertainty.
It gives rise to only a small effect.
This is verified by repeating the fit procedure under the condition that data points with $y>0.45$ are excluded from the minimisation procedure, in order to reduce the impact of the $F_L^{D(3)}$  contribution.
The number of data points is then reduced to $138$ and the results are found to be the same as those of figure~\ref{alphapom} within the statistical precision.

As illustrated in figure~\ref{alphapom}, the average $\alpha_{\PO}(0)$ value obtained in this analysis together with the absence of a  $Q^2$ dependence within the statistical precision of the measurement is in very good agreement with previous determinations in diffractive DIS~\cite{Aktas:2006hy,Aktas:2006hx,micha,Chekanov:2008fh}.
It also agrees within errors with a result obtained in diffractive photoproduction~\cite{Adloff:1997mi}.

\section{Conclusions}

A measurement of the reduced inclusive diffractive cross section $\sigma_{r}^{D(3)}(Q^2,\beta,x_{\pom})$ for the process $ep \rightarrow eXY$ with $M_Y < 1.6 \ {\rm GeV}$ and $|t| < 1 \ {\rm GeV^2}$ is presented.
New results are obtained using high statistics data taken from 1999 to 2007 by the H1 detector at HERA.
These measurements are combined with previous H1 results obtained using the same  technique for the selection of large rapidity gap events.
The combined data span more than two orders of magnitude in $Q^2$ from $3.5 \ {\rm GeV^2}$ to $1600 \ {\rm GeV^2}$ 
and cover the range $0.0017 \leq \beta \leq 0.8$ for five fixed values of $\xpom$ in the range $0.0003 \leq \xpom \leq 0.03$. 
In the best measured region for $Q^2 \ge 12$ GeV$^2$, the statistical and systematic uncertainties are at the level of $1\%$ and $5\%$, respectively, with an additional overall normalisation uncertainty of $4\%$.
%
By comparing to the proton-tagged cross section measurements, a contribution of $20\%$ of proton dissociation is found to be present in large rapidity gap data.

The combined H1 diffractive cross section measurements are compared with predictions from dipole and DPDF approaches.
A reasonable description of the data is achieved by both models. 
The predictions of the dipole model, including saturation, can describe the low $Q^2$ kinematic domain of the measurements better than the previous H1 DPDF fits.

The ratio of the diffractive to the inclusive $ep$ cross section is measured as a function of $x$, $Q^2$ and $\xpom$. 
At fixed $\xpom$ the ratio depends only weakly on $x$, except at the highest $x$ values. 
Proton PDF and dipole model predictions reproduce the behaviour of the ratio. 
This result  implies that the ratio of quark to gluon distributions is similar in the diffractive and inclusive process when considered at the same low $x$ value.

The $\xpom$ dependence of $\sigma_{r}^{D(3)}(Q^2,\beta,x_{\pom})$ is described using a model motivated by Regge phenomenology, in which a leading Pomeron and a sub-leading exchange contribute. 
With the high statistics of the present analysis, it is possible to test for a possible $Q^2$ dependence of the  Pomeron intercept with increased sensitivity.
The results do not exhibit any dependence on $Q^2$.
An average value of the effective Pomeron intercept over the full range in $Q^2$ can thus be obtained, which leads to
$\alpha_{\PO}(0)= 1.113 \ \pm 0.002 \ \mathrm{(exp.)}  \ 
^{+ 0.029}_{- 0.015} \ \mathrm{(model)}$.
This result is compatible with previous determinations and supports the proton vertex factorisation hypothesis.

\section*{Acknowledgements}

We are grateful to the HERA machine group whose outstanding
efforts have made this experiment possible. 
We thank the engineers and technicians for their work in constructing 
and maintaining the H1 detector, our funding agencies for financial 
support, the DESY technical staff for continual assistance and the 
DESY directorate for the hospitality which they extend to the non-DESY 
members of the collaboration.
We would like to thank C.~Marquet for helpful discussions and for providing us with the dipole model predictions.


\newpage

\begin{table} 
\begin{footnotesize} 
\begin{center}
\begin{tabular}{|c | c | c || c | c | c | c |} 
\hline
  &   &  &  &  & &  \\[-10pt]
\multicolumn{1}{|c|}{$\xpom$} & \multicolumn{1}{c|}{$Q^2$} & \multicolumn{1}{c||}{$\beta$} & \multicolumn{1}{c|}{$\xpom \sigma_r^{D(3)}$} & \multicolumn{1}{c|}{$\delta_{unc}$} & \multicolumn{1}{c|}{$\delta_{sys}$} & \multicolumn{1}{c|}{$\delta_{tot}$}\\
 & \multicolumn{1}{c|}{$[\rm{GeV^2}]$} & & & \multicolumn{1}{c|}{[\%]} & \multicolumn{1}{c|}{[\%]} & \multicolumn{1}{c|}{[\%]}  \\
 \hline\hline
 
0.0003 &    3.5 & 0.1700 & 0.02481 & 18.3 & 6.5 & 19.4 \\ 
0.0003 &    3.5 & 0.2700 & 0.02327 & 4.4 & 4.7 & 6.4 \\ 
0.0003 &    3.5 & 0.4300 & 0.03720 & 3.9 & 3.9 & 5.5 \\ 
0.0003 &    3.5 & 0.6700 & 0.04880 & 4.2 & 4.5 & 6.1 \\ 
0.0003 &    5.0 & 0.2700 & 0.03142 & 6.1 & 5.2 & 8.0 \\ 
0.0003 &    5.0 & 0.4300 & 0.04465 & 4.6 & 4.2 & 6.2 \\ 
0.0003 &    5.0 & 0.6700 & 0.05977 & 4.7 & 4.7 & 6.6 \\ 
0.0003 &    6.5 & 0.4300 & 0.05005 & 6.0 & 5.1 & 7.8 \\ 
0.0003 &    6.5 & 0.6700 & 0.06865 & 5.4 & 4.8 & 7.2 \\ 
0.0003 &    8.5 & 0.4300 & 0.03764 & 18.1 & 6.3 & 19.2 \\ 
0.0003 &    8.5 & 0.6700 & 0.06919 & 6.3 & 5.0 & 8.1 \\ 
0.0003 &   12.0 & 0.6700 & 0.06314 & 1.9 & 5.0 & 5.3 \\ 
0.0010 &    3.5 & 0.0500 & 0.01945 & 15.7 & 7.3 & 17.3 \\ 
0.0010 &    3.5 & 0.0800 & 0.02203 & 4.4 & 5.1 & 6.7 \\ 
0.0010 &    3.5 & 0.1300 & 0.02087 & 4.2 & 4.2 & 5.9 \\ 
0.0010 &    3.5 & 0.2000 & 0.02188 & 4.3 & 4.1 & 6.0 \\ 
0.0010 &    3.5 & 0.3200 & 0.02622 & 4.1 & 3.9 & 5.7 \\ 
0.0010 &    3.5 & 0.5000 & 0.02897 & 6.2 & 3.3 & 7.0 \\ 
0.0010 &    3.5 & 0.8000 & 0.04622 & 7.9 & 4.6 & 9.1 \\ 
0.0010 &    5.0 & 0.0800 & 0.02777 & 6.1 & 4.3 & 7.4 \\ 
0.0010 &    5.0 & 0.1300 & 0.02411 & 4.7 & 4.3 & 6.4 \\ 
0.0010 &    5.0 & 0.2000 & 0.02495 & 4.5 & 4.1 & 6.1 \\ 
0.0010 &    5.0 & 0.3200 & 0.03026 & 4.3 & 4.0 & 5.9 \\ 
0.0010 &    5.0 & 0.5000 & 0.03570 & 4.3 & 3.3 & 5.4 \\ 
0.0010 &    5.0 & 0.8000 & 0.04197 & 5.4 & 5.4 & 7.6 \\ 
0.0010 &    6.5 & 0.1300 & 0.02825 & 5.8 & 3.9 & 7.0 \\ 
0.0010 &    6.5 & 0.2000 & 0.03057 & 5.0 & 4.1 & 6.5 \\ 
0.0010 &    6.5 & 0.3200 & 0.03104 & 5.1 & 3.7 & 6.2 \\ 
0.0010 &    6.5 & 0.5000 & 0.03740 & 4.7 & 3.5 & 5.9 \\ 
0.0010 &    6.5 & 0.8000 & 0.05006 & 5.3 & 5.2 & 7.4 \\ 
0.0010 &    8.5 & 0.1300 & 0.03321 & 8.0 & 4.9 & 9.4 \\ 
0.0010 &    8.5 & 0.2000 & 0.03233 & 5.2 & 3.8 & 6.4 \\ 
0.0010 &    8.5 & 0.3200 & 0.03332 & 4.9 & 3.6 & 6.1 \\ 
0.0010 &    8.5 & 0.5000 & 0.03871 & 5.3 & 3.7 & 6.4 \\ 
0.0010 &    8.5 & 0.8000 & 0.04488 & 6.1 & 4.6 & 7.6 \\ 
0.0010 &   12.0 & 0.2000 & 0.03227 & 1.8 & 3.3 & 3.8 \\ 
0.0010 &   12.0 & 0.3200 & 0.03650 & 1.9 & 3.2 & 3.7 \\ 
0.0010 &   12.0 & 0.5000 & 0.04438 & 2.3 & 3.2 & 3.9 \\ 
0.0010 &   12.0 & 0.8000 & 0.05118 & 2.7 & 4.4 & 5.1 \\ 
0.0010 &   15.0 & 0.2000 & 0.04107 & 11.8 & 4.5 & 12.6 \\ 
0.0010 &   15.0 & 0.3200 & 0.03840 & 1.8 & 3.2 & 3.6 \\ 
0.0010 &   15.0 & 0.5000 & 0.04522 & 2.1 & 3.3 & 3.9 \\ 
0.0010 &   15.0 & 0.8000 & 0.04816 & 2.7 & 4.2 & 5.0 \\ 
0.0010 &   20.0 & 0.3200 & 0.03892 & 1.9 & 3.2 & 3.8 \\ 
0.0010 &   20.0 & 0.5000 & 0.04528 & 2.1 & 3.2 & 3.9 \\ 
0.0010 &   20.0 & 0.8000 & 0.04510 & 2.7 & 4.1 & 4.9 \\

\hline 
\end{tabular}
\end{center} 
\end{footnotesize} 
\caption{The reduced diffractive cross section from combined H1 LRG data $\xpom \sigma_r^{D(3)}$ quoted at fixed
$Q^2$, $\beta$ and $\xpom$ (columns 1--4). The uncorrelated and statistical ($\delta_{unc}$), correlated systematic ($\delta_{sys}$), and total ($\delta_{tot}$) uncertainties are given in columns $5$ to $7$.
All uncertainties are given in per cent. 
The overall normalisation uncertainty of $4$\% is not included.
The table continues on the next pages.}
\label{tab:data}
\end{table} 

\clearpage

\begin{table} 
\begin{center}
\begin{footnotesize} 
\begin{tabular}{|c | c | c || c | c | c | c |} 
\hline 
  &   &  &  &  & &  \\[-10pt]
\multicolumn{1}{|c|}{$\xpom$} & \multicolumn{1}{c|}{$Q^2$} & \multicolumn{1}{c||}{$\beta$} & \multicolumn{1}{c|}{$\xpom \sigma_r^{D(3)}$} & \multicolumn{1}{c|}{$\delta_{unc}$} & \multicolumn{1}{c|}{$\delta_{sys}$} & \multicolumn{1}{c|}{$\delta_{tot}$}\\ 
 & \multicolumn{1}{c|}{$[\rm{GeV^2}]$} & & & \multicolumn{1}{c|}{[\%]} & \multicolumn{1}{c|}{[\%]} & \multicolumn{1}{c|}{[\%]}  \\
 \hline\hline
 
0.0010 &   25.0 & 0.3200 & 0.05186 & 24.0 & 4.6 & 24.4 \\ 
0.0010 &   25.0 & 0.5000 & 0.04764 & 2.0 & 3.4 & 4.0 \\ 
0.0010 &   25.0 & 0.8000 & 0.04499 & 2.7 & 3.9 & 4.8 \\ 
0.0010 &   35.0 & 0.5000 & 0.04718 & 2.6 & 3.6 & 4.4 \\ 
0.0010 &   35.0 & 0.8000 & 0.04344 & 3.1 & 4.0 & 5.1 \\ 
0.0010 &   45.0 & 0.8000 & 0.04048 & 3.7 & 4.5 & 5.8 \\ 
0.0030 &    3.5 & 0.0170 & 0.01604 & 17.3 & 7.1 & 18.6 \\ 
0.0030 &    3.5 & 0.0270 & 0.01785 & 5.0 & 4.6 & 6.8 \\ 
0.0030 &    3.5 & 0.0430 & 0.01585 & 4.9 & 3.4 & 6.0 \\ 
0.0030 &    3.5 & 0.0670 & 0.01758 & 4.8 & 3.6 & 6.0 \\ 
0.0030 &    3.5 & 0.1100 & 0.01841 & 7.3 & 3.3 & 8.0 \\ 
0.0030 &    3.5 & 0.1700 & 0.01678 & 7.7 & 3.3 & 8.4 \\ 
0.0030 &    3.5 & 0.2700 & 0.02215 & 9.0 & 4.6 & 10.1 \\ 
0.0030 &    5.0 & 0.0270 & 0.02143 & 7.0 & 4.7 & 8.4 \\ 
0.0030 &    5.0 & 0.0430 & 0.02163 & 5.3 & 3.9 & 6.5 \\ 
0.0030 &    5.0 & 0.0670 & 0.01994 & 5.1 & 3.5 & 6.2 \\ 
0.0030 &    5.0 & 0.1100 & 0.01834 & 5.1 & 3.4 & 6.1 \\ 
0.0030 &    5.0 & 0.1700 & 0.02068 & 4.9 & 3.5 & 6.0 \\ 
0.0030 &    5.0 & 0.2700 & 0.02472 & 4.6 & 3.5 & 5.8 \\ 
0.0030 &    5.0 & 0.4300 & 0.02922 & 11.2 & 5.6 & 12.5 \\ 
0.0030 &    6.5 & 0.0430 & 0.02452 & 6.4 & 3.8 & 7.5 \\ 
0.0030 &    6.5 & 0.0670 & 0.02060 & 5.9 & 3.4 & 6.8 \\ 
0.0030 &    6.5 & 0.1100 & 0.02079 & 5.7 & 3.4 & 6.7 \\ 
0.0030 &    6.5 & 0.1700 & 0.01880 & 5.4 & 3.3 & 6.3 \\ 
0.0030 &    6.5 & 0.2700 & 0.02256 & 5.2 & 3.4 & 6.2 \\ 
0.0030 &    6.5 & 0.4300 & 0.02785 & 5.2 & 3.5 & 6.3 \\ 
0.0030 &    8.5 & 0.0430 & 0.02783 & 9.2 & 4.2 & 10.1 \\ 
0.0030 &    8.5 & 0.0670 & 0.02460 & 6.1 & 3.4 & 6.9 \\ 
0.0030 &    8.5 & 0.1100 & 0.02097 & 5.9 & 3.5 & 6.8 \\ 
0.0030 &    8.5 & 0.1700 & 0.02308 & 5.3 & 3.3 & 6.3 \\ 
0.0030 &    8.5 & 0.2700 & 0.02265 & 5.1 & 3.3 & 6.1 \\ 
0.0030 &    8.5 & 0.4300 & 0.03263 & 4.9 & 3.3 & 5.9 \\ 
0.0030 &    8.5 & 0.6700 & 0.04341 & 6.5 & 3.7 & 7.4 \\ 
0.0030 &   12.0 & 0.0670 & 0.02712 & 2.3 & 3.4 & 4.1 \\ 
0.0030 &   12.0 & 0.1100 & 0.02698 & 2.2 & 3.3 & 3.9 \\ 
0.0030 &   12.0 & 0.1700 & 0.02655 & 2.2 & 3.1 & 3.8 \\ 
0.0030 &   12.0 & 0.2700 & 0.02751 & 2.3 & 3.2 & 4.0 \\ 
0.0030 &   12.0 & 0.4300 & 0.03388 & 2.3 & 3.2 & 4.0 \\ 
0.0030 &   12.0 & 0.6700 & 0.04193 & 2.6 & 3.3 & 4.2 \\ 
0.0030 &   15.0 & 0.0670 & 0.03764 & 14.2 & 4.7 & 14.9 \\ 
0.0030 &   15.0 & 0.1100 & 0.02780 & 2.1 & 3.4 & 3.9 \\ 
0.0030 &   15.0 & 0.1700 & 0.02732 & 2.1 & 3.2 & 3.8 \\ 
0.0030 &   15.0 & 0.2700 & 0.02903 & 2.0 & 3.1 & 3.7 \\ 
0.0030 &   15.0 & 0.4300 & 0.03449 & 2.2 & 3.1 & 3.8 \\ 
0.0030 &   15.0 & 0.6700 & 0.04031 & 2.5 & 3.2 & 4.1 \\ 
0.0030 &   20.0 & 0.1100 & 0.02754 & 2.3 & 3.3 & 4.1 \\

\hline 
\end{tabular}
\end{footnotesize} 
\end{center} 
\caption{The reduced diffractive cross section from combined H1 LRG data $\xpom \sigma_r^{D(3)}$ quoted at fixed
$Q^2$, $\beta$ and $\xpom$, continued from table~\ref{tab:data}.}
\end{table}

\begin{table}
\begin{center}
\begin{footnotesize} 
\begin{tabular}{|c | c | c || c | c | c | c |} 
\hline 
  &   &  &  &  & &  \\[-10pt]
\multicolumn{1}{|c|}{$\xpom$} & \multicolumn{1}{c|}{$Q^2$} & \multicolumn{1}{c||}{$\beta$} & \multicolumn{1}{c|}{$\xpom \sigma_r^{D(3)}$} & \multicolumn{1}{c|}{$\delta_{unc}$} & \multicolumn{1}{c|}{$\delta_{sys}$} & \multicolumn{1}{c|}{$\delta_{tot}$}\\ 
 & \multicolumn{1}{c|}{$[\rm{GeV^2}]$} & & & \multicolumn{1}{c|}{[\%]} & \multicolumn{1}{c|}{[\%]} & \multicolumn{1}{c|}{[\%]}  \\
 \hline\hline
 
0.0030 &   20.0 & 0.1700 & 0.02996 & 2.0 & 3.2 & 3.8 \\ 
0.0030 &   20.0 & 0.2700 & 0.03194 & 1.9 & 3.0 & 3.6 \\ 
0.0030 &   20.0 & 0.4300 & 0.03618 & 2.0 & 3.1 & 3.7 \\ 
0.0030 &   20.0 & 0.6700 & 0.03927 & 2.4 & 3.3 & 4.1 \\ 
0.0030 &   25.0 & 0.1100 & 0.03645 & 20.2 & 5.4 & 20.9 \\ 
0.0030 &   25.0 & 0.1700 & 0.03156 & 2.1 & 3.0 & 3.7 \\ 
0.0030 &   25.0 & 0.2700 & 0.03205 & 1.9 & 3.0 & 3.6 \\ 
0.0030 &   25.0 & 0.4300 & 0.03706 & 2.0 & 3.0 & 3.6 \\ 
0.0030 &   25.0 & 0.6700 & 0.03909 & 2.4 & 3.4 & 4.2 \\ 
0.0030 &   35.0 & 0.1700 & 0.03132 & 2.5 & 3.3 & 4.1 \\ 
0.0030 &   35.0 & 0.2700 & 0.03330 & 2.0 & 2.9 & 3.6 \\ 
0.0030 &   35.0 & 0.4300 & 0.03691 & 2.1 & 3.1 & 3.7 \\ 
0.0030 &   35.0 & 0.6700 & 0.03975 & 2.5 & 3.7 & 4.4 \\ 
0.0030 &   45.0 & 0.2700 & 0.03306 & 2.4 & 3.0 & 3.9 \\ 
0.0030 &   45.0 & 0.4300 & 0.03872 & 2.3 & 3.0 & 3.8 \\ 
0.0030 &   45.0 & 0.6700 & 0.03844 & 2.8 & 3.7 & 4.6 \\ 
0.0030 &   60.0 & 0.4300 & 0.03776 & 2.7 & 3.1 & 4.1 \\ 
0.0030 &   60.0 & 0.6700 & 0.03728 & 3.1 & 3.7 & 4.8 \\ 
0.0030 &   90.0 & 0.6700 & 0.03532 & 5.4 & 4.4 & 6.9 \\ 
0.0100 &    3.5 & 0.0050 & 0.02678 & 16.0 & 6.1 & 17.1 \\ 
0.0100 &    3.5 & 0.0080 & 0.02007 & 6.7 & 4.3 & 7.9 \\ 
0.0100 &    3.5 & 0.0130 & 0.01938 & 6.8 & 3.9 & 7.9 \\ 
0.0100 &    3.5 & 0.0200 & 0.01632 & 6.3 & 3.5 & 7.2 \\ 
0.0100 &    3.5 & 0.0320 & 0.01795 & 9.3 & 4.0 & 10.1 \\ 
0.0100 &    3.5 & 0.0500 & 0.01554 & 9.8 & 3.7 & 10.5 \\ 
0.0100 &    3.5 & 0.0800 & 0.01729 & 11.0 & 4.4 & 11.8 \\ 
0.0100 &    5.0 & 0.0080 & 0.02647 & 7.5 & 4.8 & 8.9 \\ 
0.0100 &    5.0 & 0.0130 & 0.02361 & 6.7 & 4.0 & 7.8 \\ 
0.0100 &    5.0 & 0.0200 & 0.02137 & 6.4 & 3.6 & 7.4 \\ 
0.0100 &    5.0 & 0.0320 & 0.02000 & 6.3 & 3.5 & 7.2 \\ 
0.0100 &    5.0 & 0.0500 & 0.01922 & 6.3 & 3.5 & 7.2 \\ 
0.0100 &    5.0 & 0.0800 & 0.01657 & 6.9 & 3.9 & 8.0 \\ 
0.0100 &    6.5 & 0.0130 & 0.02516 & 7.2 & 3.8 & 8.1 \\ 
0.0100 &    6.5 & 0.0200 & 0.02356 & 6.9 & 3.3 & 7.7 \\ 
0.0100 &    6.5 & 0.0320 & 0.02270 & 6.4 & 3.3 & 7.2 \\ 
0.0100 &    6.5 & 0.0500 & 0.02205 & 6.8 & 3.5 & 7.6 \\ 
0.0100 &    6.5 & 0.0800 & 0.01938 & 5.9 & 3.6 & 6.9 \\ 
0.0100 &    6.5 & 0.1300 & 0.01757 & 6.7 & 3.4 & 7.5 \\ 
0.0100 &    8.5 & 0.0130 & 0.03654 & 9.2 & 4.0 & 10.0 \\ 
0.0100 &    8.5 & 0.0200 & 0.03174 & 6.2 & 3.8 & 7.3 \\ 
0.0100 &    8.5 & 0.0320 & 0.03085 & 5.8 & 3.4 & 6.7 \\ 
0.0100 &    8.5 & 0.0500 & 0.02431 & 6.1 & 3.2 & 6.9 \\ 
0.0100 &    8.5 & 0.0800 & 0.02142 & 5.9 & 3.5 & 6.8 \\ 
0.0100 &    8.5 & 0.1300 & 0.01919 & 6.1 & 3.6 & 7.1 \\ 
0.0100 &    8.5 & 0.2000 & 0.01961 & 7.2 & 3.3 & 7.9 \\ 
0.0100 &   12.0 & 0.0200 & 0.03014 & 3.9 & 3.3 & 5.1 \\

\hline 
\end{tabular}
\end{footnotesize} 
\end{center} 
\caption{The reduced diffractive cross section from combined H1 LRG data $\xpom \sigma_r^{D(3)}$ quoted at fixed
$Q^2$, $\beta$ and $\xpom$, continued from table~\ref{tab:data}.}
\end{table}

\begin{table}
\begin{center}
\begin{footnotesize} 
\begin{tabular}{|c | c | c || c | c | c | c |} 
\hline 
  &   &  &  &  & &  \\[-10pt]
\multicolumn{1}{|c|}{$\xpom$} & \multicolumn{1}{c|}{$Q^2$} & \multicolumn{1}{c||}{$\beta$} & \multicolumn{1}{c|}{$\xpom \sigma_r^{D(3)}$} & \multicolumn{1}{c|}{$\delta_{unc}$} & \multicolumn{1}{c|}{$\delta_{sys}$} & \multicolumn{1}{c|}{$\delta_{tot}$}\\ 
 & \multicolumn{1}{c|}{$[\rm{GeV^2}]$} & & & \multicolumn{1}{c|}{[\%]} & \multicolumn{1}{c|}{[\%]} & \multicolumn{1}{c|}{[\%]}  \\
 \hline\hline
 
0.0100 &   12.0 & 0.0320 & 0.02966 & 3.8 & 3.2 & 5.0 \\ 
0.0100 &   12.0 & 0.0500 & 0.02732 & 3.7 & 3.2 & 4.9 \\ 
0.0100 &   12.0 & 0.0800 & 0.02545 & 3.6 & 3.2 & 4.8 \\ 
0.0100 &   12.0 & 0.1300 & 0.02165 & 3.8 & 3.2 & 4.9 \\ 
0.0100 &   12.0 & 0.2000 & 0.02283 & 4.2 & 3.1 & 5.2 \\ 
0.0100 &   12.0 & 0.3200 & 0.02420 & 5.1 & 3.1 & 6.0 \\ 
0.0100 &   15.0 & 0.0200 & 0.03639 & 14.5 & 5.8 & 15.6 \\ 
0.0100 &   15.0 & 0.0320 & 0.03226 & 3.2 & 3.4 & 4.7 \\ 
0.0100 &   15.0 & 0.0500 & 0.03067 & 3.1 & 3.3 & 4.6 \\ 
0.0100 &   15.0 & 0.0800 & 0.02573 & 3.2 & 3.3 & 4.6 \\ 
0.0100 &   15.0 & 0.1300 & 0.02381 & 3.0 & 3.3 & 4.5 \\ 
0.0100 &   15.0 & 0.2000 & 0.02299 & 3.0 & 3.3 & 4.5 \\ 
0.0100 &   15.0 & 0.3200 & 0.02456 & 3.3 & 3.1 & 4.5 \\ 
0.0100 &   20.0 & 0.0320 & 0.03445 & 4.0 & 3.3 & 5.2 \\ 
0.0100 &   20.0 & 0.0500 & 0.03209 & 3.3 & 3.4 & 4.8 \\ 
0.0100 &   20.0 & 0.0800 & 0.02971 & 3.5 & 3.3 & 4.8 \\ 
0.0100 &   20.0 & 0.1300 & 0.02658 & 3.1 & 3.2 & 4.5 \\ 
0.0100 &   20.0 & 0.2000 & 0.02542 & 3.4 & 3.2 & 4.7 \\ 
0.0100 &   20.0 & 0.3200 & 0.02663 & 3.1 & 3.2 & 4.4 \\ 
0.0100 &   20.0 & 0.5000 & 0.02870 & 3.7 & 3.2 & 4.8 \\ 
0.0100 &   25.0 & 0.0320 & 0.03306 & 19.8 & 6.4 & 20.8 \\ 
0.0100 &   25.0 & 0.0500 & 0.03307 & 3.2 & 3.5 & 4.8 \\ 
0.0100 &   25.0 & 0.0800 & 0.03202 & 3.2 & 3.4 & 4.7 \\ 
0.0100 &   25.0 & 0.1300 & 0.02889 & 3.2 & 3.4 & 4.6 \\ 
0.0100 &   25.0 & 0.2000 & 0.02686 & 3.0 & 3.3 & 4.5 \\ 
0.0100 &   25.0 & 0.3200 & 0.02769 & 3.1 & 3.4 & 4.6 \\ 
0.0100 &   25.0 & 0.5000 & 0.03028 & 3.4 & 3.3 & 4.7 \\ 
0.0100 &   25.0 & 0.8000 & 0.02928 & 7.0 & 3.8 & 7.9 \\ 
0.0100 &   35.0 & 0.0500 & 0.03551 & 4.1 & 3.5 & 5.3 \\ 
0.0100 &   35.0 & 0.0800 & 0.03243 & 3.8 & 3.3 & 5.0 \\ 
0.0100 &   35.0 & 0.1300 & 0.03161 & 3.2 & 3.3 & 4.6 \\ 
0.0100 &   35.0 & 0.2000 & 0.02963 & 3.3 & 3.1 & 4.5 \\ 
0.0100 &   35.0 & 0.3200 & 0.02729 & 3.2 & 3.7 & 4.9 \\ 
0.0100 &   35.0 & 0.5000 & 0.03171 & 3.5 & 3.1 & 4.7 \\ 
0.0100 &   35.0 & 0.8000 & 0.02840 & 4.3 & 3.5 & 5.5 \\ 
0.0100 &   45.0 & 0.0800 & 0.03368 & 4.1 & 3.3 & 5.3 \\ 
0.0100 &   45.0 & 0.1300 & 0.03212 & 3.4 & 3.2 & 4.6 \\ 
0.0100 &   45.0 & 0.2000 & 0.02994 & 3.4 & 3.2 & 4.7 \\ 
0.0100 &   45.0 & 0.3200 & 0.02910 & 3.3 & 3.5 & 4.8 \\ 
0.0100 &   45.0 & 0.5000 & 0.03255 & 3.7 & 3.0 & 4.8 \\ 
0.0100 &   45.0 & 0.8000 & 0.02606 & 4.5 & 3.5 & 5.7 \\ 
0.0100 &   60.0 & 0.1300 & 0.03316 & 4.1 & 3.1 & 5.2 \\ 
0.0100 &   60.0 & 0.2000 & 0.03013 & 3.3 & 3.3 & 4.7 \\ 
0.0100 &   60.0 & 0.3200 & 0.03138 & 3.4 & 3.1 & 4.6 \\ 
0.0100 &   60.0 & 0.5000 & 0.03225 & 3.6 & 3.7 & 5.2 \\ 
0.0100 &   60.0 & 0.8000 & 0.02516 & 4.0 & 3.7 & 5.4 \\

\hline 
\end{tabular}
\end{footnotesize} 
\end{center} 
\caption{The reduced diffractive cross section from combined H1 LRG data $\xpom \sigma_r^{D(3)}$ quoted at fixed
$Q^2$, $\beta$ and $\xpom$, continued from table~\ref{tab:data}.}
\end{table}

\begin{table}
\begin{center}
\begin{footnotesize} 
\begin{tabular}{|c | c | c || c | c | c | c |} 
\hline 
  &   &  &  &  & &  \\[-10pt]
\multicolumn{1}{|c|}{$\xpom$} & \multicolumn{1}{c|}{$Q^2$} & \multicolumn{1}{c||}{$\beta$} & \multicolumn{1}{c|}{$\xpom \sigma_r^{D(3)}$} & \multicolumn{1}{c|}{$\delta_{unc}$} & \multicolumn{1}{c|}{$\delta_{sys}$} & \multicolumn{1}{c|}{$\delta_{tot}$}\\ 
 & \multicolumn{1}{c|}{$[\rm{GeV^2}]$} & & & \multicolumn{1}{c|}{[\%]} & \multicolumn{1}{c|}{[\%]} & \multicolumn{1}{c|}{[\%]}  \\
 \hline\hline
 
0.0100 &   90.0 & 0.2000 & 0.03061 & 5.0 & 3.5 & 6.2 \\ 
0.0100 &   90.0 & 0.3200 & 0.03095 & 4.3 & 3.1 & 5.2 \\ 
0.0100 &   90.0 & 0.5000 & 0.03039 & 3.8 & 3.3 & 5.1 \\ 
0.0100 &   90.0 & 0.8000 & 0.02396 & 4.3 & 3.6 & 5.7 \\ 
0.0100 &  200.0 & 0.3200 & 0.03210 & 6.8 & 8.5 & 10.9 \\ 
0.0100 &  200.0 & 0.5000 & 0.03150 & 6.4 & 8.8 & 10.9 \\ 
0.0100 &  200.0 & 0.8000 & 0.02110 & 8.7 & 8.0 & 11.8 \\ 
0.0100 &  400.0 & 0.8000 & 0.01960 & 13.8 & 9.6 & 16.7 \\ 
0.0300 &    3.5 & 0.0017 & 0.01919 & 29.3 & 8.7 & 30.6 \\ 
0.0300 &    3.5 & 0.0027 & 0.02575 & 18.0 & 8.6 & 19.9 \\ 
0.0300 &    3.5 & 0.0043 & 0.02418 & 17.0 & 7.8 & 18.7 \\ 
0.0300 &    3.5 & 0.0067 & 0.02030 & 16.9 & 6.9 & 18.2 \\ 
0.0300 &    3.5 & 0.0110 & 0.01811 & 17.6 & 6.8 & 18.9 \\ 
0.0300 &    5.0 & 0.0027 & 0.03776 & 21.1 & 14.3 & 25.5 \\ 
0.0300 &    5.0 & 0.0043 & 0.03206 & 17.8 & 6.3 & 18.9 \\ 
0.0300 &    5.0 & 0.0067 & 0.02984 & 16.2 & 7.1 & 17.7 \\ 
0.0300 &    5.0 & 0.0110 & 0.02269 & 17.7 & 6.4 & 18.9 \\ 
0.0300 &    5.0 & 0.0170 & 0.02157 & 16.7 & 7.3 & 18.2 \\ 
0.0300 &    6.5 & 0.0027 & 0.04277 & 34.1 & 8.7 & 35.2 \\ 
0.0300 &    6.5 & 0.0043 & 0.02261 & 18.4 & 7.9 & 20.1 \\ 
0.0300 &    6.5 & 0.0067 & 0.02536 & 17.3 & 7.0 & 18.6 \\ 
0.0300 &    6.5 & 0.0110 & 0.02534 & 17.4 & 7.0 & 18.7 \\ 
0.0300 &    6.5 & 0.0170 & 0.02571 & 17.0 & 5.5 & 17.9 \\ 
0.0300 &    6.5 & 0.0270 & 0.02512 & 16.3 & 6.6 & 17.6 \\ 
0.0300 &    6.5 & 0.0430 & 0.02256 & 16.8 & 6.1 & 17.9 \\ 
0.0300 &    8.5 & 0.0043 & 0.03435 & 23.1 & 8.8 & 24.7 \\ 
0.0300 &    8.5 & 0.0067 & 0.02474 & 18.6 & 5.1 & 19.3 \\ 
0.0300 &    8.5 & 0.0110 & 0.03042 & 16.1 & 5.7 & 17.1 \\ 
0.0300 &    8.5 & 0.0170 & 0.02617 & 15.8 & 6.3 & 17.0 \\ 
0.0300 &    8.5 & 0.0270 & 0.02631 & 15.3 & 6.4 & 16.6 \\ 
0.0300 &    8.5 & 0.0430 & 0.02782 & 17.1 & 6.1 & 18.1 \\ 
0.0300 &   12.0 & 0.0067 & 0.03331 & 22.0 & 5.8 & 22.7 \\ 
0.0300 &   12.0 & 0.0110 & 0.03641 & 16.7 & 4.9 & 17.4 \\ 
0.0300 &   12.0 & 0.0170 & 0.03224 & 16.3 & 6.7 & 17.6 \\ 
0.0300 &   12.0 & 0.0270 & 0.03637 & 16.1 & 6.5 & 17.4 \\ 
0.0300 &   12.0 & 0.0430 & 0.02906 & 17.5 & 5.5 & 18.4 \\ 
0.0300 &   12.0 & 0.0670 & 0.02413 & 17.6 & 5.2 & 18.3 \\ 
0.0300 &   15.0 & 0.0067 & 0.04792 & 19.4 & 6.4 & 20.4 \\ 
0.0300 &   15.0 & 0.0110 & 0.03531 & 13.7 & 6.6 & 15.2 \\ 
0.0300 &   15.0 & 0.0170 & 0.03527 & 12.6 & 6.4 & 14.1 \\ 
0.0300 &   15.0 & 0.0270 & 0.03085 & 13.3 & 5.9 & 14.5 \\ 
0.0300 &   15.0 & 0.0430 & 0.02592 & 13.4 & 7.1 & 15.2 \\ 
0.0300 &   15.0 & 0.0670 & 0.02366 & 13.3 & 5.9 & 14.5 \\ 
0.0300 &   15.0 & 0.1100 & 0.02278 & 13.7 & 6.4 & 15.2 \\ 
0.0300 &   20.0 & 0.0110 & 0.03178 & 15.6 & 7.2 & 17.2 \\ 
0.0300 &   20.0 & 0.0170 & 0.03851 & 14.0 & 6.2 & 15.4 \\

\hline 
\end{tabular}
\end{footnotesize} 
\end{center} 
\caption{The reduced diffractive cross section from combined H1 LRG data $\xpom \sigma_r^{D(3)}$ quoted at fixed
$Q^2$, $\beta$ and $\xpom$, continued from table~\ref{tab:data}.}
\end{table}

\begin{table}
\begin{center}
\begin{footnotesize} 
\begin{tabular}{|c | c | c || c | c | c | c |} 
\hline 
  &   &  &  &  & &  \\[-10pt]
\multicolumn{1}{|c|}{$\xpom$} & \multicolumn{1}{c|}{$Q^2$} & \multicolumn{1}{c||}{$\beta$} & \multicolumn{1}{c|}{$\xpom \sigma_r^{D(3)}$} & \multicolumn{1}{c|}{$\delta_{unc}$} & \multicolumn{1}{c|}{$\delta_{sys}$} & \multicolumn{1}{c|}{$\delta_{tot}$}\\ 
 & \multicolumn{1}{c|}{$[\rm{GeV^2}]$} & & & \multicolumn{1}{c|}{[\%]} & \multicolumn{1}{c|}{[\%]} & \multicolumn{1}{c|}{[\%]}  \\
 \hline\hline
 
0.0300 &   20.0 & 0.0270 & 0.03118 & 12.9 & 5.5 & 14.1 \\ 
0.0300 &   20.0 & 0.0430 & 0.02917 & 12.9 & 5.5 & 14.0 \\ 
0.0300 &   20.0 & 0.0670 & 0.02773 & 13.0 & 5.5 & 14.1 \\ 
0.0300 &   20.0 & 0.1100 & 0.02288 & 13.5 & 5.8 & 14.7 \\ 
0.0300 &   25.0 & 0.0110 & 0.03729 & 28.0 & 7.4 & 29.0 \\ 
0.0300 &   25.0 & 0.0170 & 0.03875 & 14.3 & 6.3 & 15.6 \\ 
0.0300 &   25.0 & 0.0270 & 0.03755 & 13.1 & 5.6 & 14.3 \\ 
0.0300 &   25.0 & 0.0430 & 0.02978 & 13.1 & 4.7 & 14.0 \\ 
0.0300 &   25.0 & 0.0670 & 0.02655 & 13.8 & 6.6 & 15.3 \\ 
0.0300 &   25.0 & 0.1100 & 0.02491 & 13.1 & 6.4 & 14.5 \\ 
0.0300 &   25.0 & 0.1700 & 0.02562 & 13.3 & 6.4 & 14.7 \\ 
0.0300 &   35.0 & 0.0170 & 0.05337 & 18.9 & 6.1 & 19.9 \\ 
0.0300 &   35.0 & 0.0270 & 0.04213 & 13.8 & 5.0 & 14.6 \\ 
0.0300 &   35.0 & 0.0430 & 0.04063 & 14.0 & 4.6 & 14.7 \\ 
0.0300 &   35.0 & 0.0670 & 0.03063 & 13.6 & 6.0 & 14.8 \\ 
0.0300 &   35.0 & 0.1100 & 0.02992 & 13.4 & 6.3 & 14.8 \\ 
0.0300 &   35.0 & 0.1700 & 0.02493 & 13.8 & 6.2 & 15.1 \\ 
0.0300 &   35.0 & 0.2700 & 0.02840 & 13.5 & 6.6 & 15.1 \\ 
0.0300 &   45.0 & 0.0270 & 0.05064 & 17.0 & 4.9 & 17.6 \\ 
0.0300 &   45.0 & 0.0430 & 0.04048 & 14.3 & 4.4 & 15.0 \\ 
0.0300 &   45.0 & 0.0670 & 0.03804 & 15.4 & 6.1 & 16.5 \\ 
0.0300 &   45.0 & 0.1100 & 0.02427 & 14.6 & 6.6 & 16.0 \\ 
0.0300 &   45.0 & 0.1700 & 0.02521 & 14.2 & 7.1 & 15.9 \\ 
0.0300 &   45.0 & 0.2700 & 0.02092 & 14.4 & 6.5 & 15.8 \\ 
0.0300 &   60.0 & 0.0430 & 0.03900 & 17.7 & 5.7 & 18.6 \\ 
0.0300 &   60.0 & 0.0670 & 0.03913 & 14.7 & 5.2 & 15.6 \\ 
0.0300 &   60.0 & 0.1100 & 0.02613 & 14.6 & 5.5 & 15.6 \\ 
0.0300 &   60.0 & 0.1700 & 0.02548 & 14.3 & 8.2 & 16.5 \\ 
0.0300 &   60.0 & 0.2700 & 0.02165 & 18.2 & 8.5 & 20.1 \\ 
0.0300 &   60.0 & 0.4300 & 0.02698 & 14.7 & 8.7 & 17.1 \\ 
0.0300 &   90.0 & 0.0670 & 0.03286 & 39.0 & 6.9 & 39.6 \\ 
0.0300 &   90.0 & 0.1100 & 0.03379 & 18.5 & 4.5 & 19.1 \\ 
0.0300 &   90.0 & 0.1700 & 0.03622 & 15.6 & 6.0 & 16.7 \\ 
0.0300 &   90.0 & 0.2700 & 0.02668 & 15.4 & 5.5 & 16.4 \\ 
0.0300 &   90.0 & 0.4300 & 0.03214 & 16.1 & 5.7 & 17.1 \\ 
0.0300 &   90.0 & 0.6700 & 0.02818 & 24.5 & 7.9 & 25.8 \\ 
0.0300 &  200.0 & 0.1100 & 0.03610 & 12.5 & 9.8 & 15.9 \\ 
0.0300 &  200.0 & 0.1700 & 0.03310 & 12.1 & 9.5 & 15.4 \\ 
0.0300 &  200.0 & 0.2700 & 0.02830 & 12.3 & 8.4 & 14.9 \\ 
0.0300 &  200.0 & 0.4300 & 0.03090 & 12.4 & 8.2 & 14.9 \\ 
0.0300 &  200.0 & 0.6700 & 0.02970 & 13.2 & 10.1 & 16.6 \\ 
0.0300 &  400.0 & 0.2700 & 0.03220 & 13.5 & 9.9 & 16.7 \\ 
0.0300 &  400.0 & 0.4300 & 0.02930 & 13.1 & 8.2 & 15.4 \\ 
0.0300 &  400.0 & 0.6700 & 0.02890 & 13.7 & 10.2 & 17.0 \\ 
0.0300 &  800.0 & 0.4300 & 0.03910 & 17.2 & 10.3 & 20.1 \\ 
0.0300 &  800.0 & 0.6700 & 0.02280 & 18.3 & 11.6 & 21.6 \\ 
0.0300 & 1600.0 & 0.6700 & 0.02140 & 30.0 & 12.8 & 32.6 \\

\hline 
\end{tabular}
\end{footnotesize} 
\end{center} 
\caption{The reduced diffractive cross section from combined H1 LRG data $\xpom \sigma_r^{D(3)}$ quoted at fixed
$Q^2$, $\beta$ and $\xpom$, continued from table~\ref{tab:data}.}
\label{tab:data_end}
\end{table} 


\newpage

\end{document}